\documentclass[aps,nofootinbib,amssymb,twocolumn,superscriptaddress,longbibliography,10pt]{revtex4-1}
\usepackage[utf8]{inputenc}
\usepackage{color}
\usepackage{times}
\usepackage{array}
\usepackage{verbatim}
\usepackage{multirow}
\usepackage{amsmath}
\usepackage{amssymb}
\usepackage{dsfont}
\usepackage{graphicx}
\usepackage{esint}
\usepackage[unicode=true,
 bookmarks=true,bookmarksnumbered=true,bookmarksopen=false,
 breaklinks=true,pdfborder={0 0 0},backref=false,colorlinks=true]
 {hyperref}
\usepackage{enumitem}
\usepackage{titlesec}
\usepackage{cleveref}
\usepackage{longtable}
\usepackage{bbm}
\usepackage[normalem]{ulem}
\usepackage{listings}
\usepackage{xcolor,colortbl}

\hypersetup{
citecolor={blue},
urlcolor={blue}
}

\usepackage{cleveref}
\crefname{appendix}{Appendix}{Appendices}
\crefname{equation}{Eq.}{Eqs.}
\crefname{figure}{Fig.}{Figs.}
\crefname{table}{Table}{Tables}
\crefname{section}{Section}{Sections}
\crefname{mythe}{Theorem}{Theorems}
\crefname{mydef}{Definition}{Definitions}

\renewcommand{\paragraph}[1]{\vspace{0.2cm}{\bf \textit{#1}}}

\usepackage{simpler-wick}
\allowdisplaybreaks[1] 

\newcommand{\td}{\widetilde}

\def\pare#1{\left( #1 \right)}
\def\brak#1{\left[#1\right]}

\def\inn#1{\langle #1 \rangle}

\def\up{\uparrow}
\def\down{\downarrow}

\def\kk{\mathbf{k}}

\def\qq{\mathbf{q}}
\def\rr{\mathbf{r}}

\def\GG{\mathbf{G}}
\def\RR{\mathbf{R}}

\def\hH{\hat{H}}

\def\ii{\mathrm{i\:\!}}  

\def\ba#1\ea{\begin{align}#1\end{align}}

\let\vec\mathbf

\newcommand{\vk}{\vec{k}}


\begin{document}
\title{Projected and Solvable  Topological Heavy Fermion Model of Twisted Bilayer Graphene}

\author{Haoyu Hu}
\affiliation{Department of Physics, Princeton University, Princeton, NJ 08544, USA}

\author{Zhi-Da Song}
\affiliation{International Center for Quantum Materials, School of Physics, Peking University, Beijing 100871, China}
\affiliation{Hefei National Laboratory, Hefei 230088, China}
\affiliation{Collaborative Innovation Center of Quantum Matter, Beijing 100871, China}

\author{B.~Andrei Bernevig}
\affiliation{Department of Physics, Princeton University, Princeton, NJ 08544, USA}
\affiliation{Donostia International Physics Center (DIPC), Paseo Manuel de Lardizábal. 20018, San Sebastián, Spain}
\affiliation{IKERBASQUE, Basque Foundation for Science, 48013 Bilbao, Spain}

\begin{abstract}
We investigate the topological heavy-fermion (THF) model of magic-angle twisted bilayer graphene (MATBG) in the projected limit, where only the flat bands are present in the low-energy spectrum. Such limit has been previously analyzed in momentum-space Bistritzer-MacDonald-type continuum models, but not in a real-space formalism. 
In this regime, the Hubbard interaction ($U_1$) of the $f$-electrons is larger than the bandwidth ($2M$) of the flat bands but smaller than the gap ($\gamma$) between the flat and remote bands.
In the THF model, concentrated charge (in real space) and concentrated Berry curvature (in momentum space) are respectively realized by exponentially localized $f$-orbitals and itinerant Dirac $c$-electrons. Local moments naturally arise from $f$-orbitals. 
Hybridizing the $f$-electrons with $c$-electrons produces power-law tails of the flat-band Wannier functions, raising the question of relevance of the local moment picture in the projected $U_1\ll \gamma$ limit. 
Nonetheless, we find that the local moments remain stable as long as $U_1 \gg \Delta(\omega)$ for $|\omega|\lesssim U_1$, where $\Delta(\omega)\sim \gamma^2 N(\omega)$ is the hybridization function seen by each $f$-site, and $N(\omega)$ is the density of states of the Dirac $c$-bands.  
Notably, the comparison between $U_1$ and $\gamma$ is irrelevant to the local moment formation if $N(\omega)$ is unknown. 
Within the framework of THF, we also derive the correlated self-energy of the flat bands using the Hubbard-I approximation, and estimate the coupling strength between the local moments. Finally, we comment that, in the regime of extremely concentrated Berry curvature, the single-particle gap between flat bands and remote bands vanishes and the interaction strength is always larger than the gap.   
\end{abstract}

\maketitle

\onecolumngrid

\section{Introduction and motivation}
The emergence of flat bands in twisted bilayer graphene gives rise to a rich variety of strongly correlated phenomena~\cite{CAO18,KER19,XIE19,SHA19,JIA19,CHO19,POL19,YAN19,LU19,STE20,SAI20,SER20,CHE20b,WON20,CHO20,NUC20,CHO21,SAI21,LIU21c,PAR21c,WU21a,CAO21,DAS21,TSC21,PIE21,STE21,CHO21a,XIE21d,DAS22,NUC23,YU23c}, including unconventional superconductivity \cite{CAO18a,YAN19,LU19,STE20,SAI20,DE21a,OH21,TIA23,DI22a,CAL22d} and exotic quantum phases~\cite{TOM19,CAO20,ZON20,LIS21,BEN21,LIA21c,ROZ21,SAI21a,LU21,HES21,DIE23,HUB22,GHA22,JAO22,PAU22,GRO22,ZHO23a}. 
Substantial theoretical efforts have been made to understand the rich physics of magic-angle twisted bilayer graphene (MATBG), focusing on constructing models \cite{LOP07,SUA10,bistritzer_moire_2011,UCH14,WIJ15,DAI16,JAI16,NAM17,EFI18,KAN18,ZOU18,PO19,LIU19a,TAR19,MOR19,LI19,FAN19,KHA19,CAR19a,CAR19,RAD19,KWA20,CAR20,TRI20,WIL20,PAR20,CAR20a,FU20,HUA20a,CAL20,WU21d,REN21,HEJ21,CAL21,BER21,BER21a,WAN21a,RAM21,SHI21,LEI21,CAO21a,SHE21,WU21c,KOS18,LED21,GUE22,DAV22,CLA22,LIN22,SAM22,KAN23b,VAF23,SHI23}, correlated states \cite{OCH18,THO18,XU18b,KOS18,PO18a,VEN18,YUA18,DOD18,PAD18,KEN18,RAD18,LIU19,HUA19,WU19,CLA19,KAN19,SEO19,DA19,ANG19,XIE20b,BUL20b,CHA20,REP20,CEA20,ZHA20,KAN20a,BUL20a,CHI20b,SOE20,CHR20,EUG20,WU20,VAF20,XIE21,KAN21,LIU21,DA21,LIU21a,THO21,KWA21a,LIA21,ZHA21,PAR21a,VAF21,KWA21b,CHE21,POT21,XIE21b,XIE21a,LED21,CHA21,KWA21,HOF22,WAG22,CHR22,BRI22,CAL22d,HON22,ZHA23a,BLA22,XIE23a,KWA23,YU23a,WAN23c,FER20,DAT23,RAI23a}, and superconductivity \cite{PhysRevB.110.045133,GUO18,YUA18,XU18,DOD18,PO18a,LIU18a,VEN18,ISO18,PEL18,KEN18,WU18,GUI18,GON19,HUA19,ROY19,WU19a,WU19,YOU19,CLA19,LIA19,HU19a,JUL20,XIE20,CHI20a,LOP20,KON20,CHR20,WAN21,KHA21,LEW21,FER21,QIN21,PHO21,CHO21d,LAK21,CHO21c,LI22c,FIS22,YU22,SCA22,CHR22,CHA22,KWA22,WAG23a,GON23,WAG23,WAN24}. 
The topological heavy-fermion (THF) model \cite{song_magic-angle_2022,calugaru_tbg_2023,shi_heavy-fermion_2022} has been proposed as a framework for comprehensively understanding MATBG.  
The THF model consists of $f$-orbitals, exponentially localized at the AA-stacking regions of the moir\'e pattern, and itinerant Dirac $c$-electrons. In this model, the strong correlations observed in MATBG are attributed to the local Coulomb repulsion of the quantum-dot-like $f$-orbitals, while the topology of the flat bands is carried by the Dirac $c$-electrons.
The $cf$ hybridization generates a topological flat band that accurately reproduces both the dispersion and interaction form factors \cite{calugaru_tbg_2023} of the continuous Bistritzer-MacDonald (BM) model \cite{bistritzer_moire_2011}. 
The local representation ($f$-orbitals) allows the use of advanced theoretical and numerical methods to study the correlation effect in MATBG \cite{hu_symmetric_2023,zhou_kondo_2024,rai_dynamical_2023,datta_heavy_2023,song_magic-angle_2022,shi_heavy-fermion_2022,chou_kondo_2023,lau_topological_2023, wang_molecular_2024, youn_hundness_2024}. 
Thus, various phases, including heavy Fermi liquid, disordered local moments arising from $f$-orbitals, RKKY-stabilized correlated insulators, unconventional superconductor and their competitions have been studied within this model using Hartree-Fock \cite{song_magic-angle_2022,shi_heavy-fermion_2022}, large-$U$ perturbation theory \cite{hu_kondo_2023}, slave-particle \cite{chou_kondo_2023,lau_topological_2023}, dynamical mean field theory (DMFT) \cite{hu_symmetric_2023,zhou_kondo_2024,rai_dynamical_2023,datta_heavy_2023,youn_hundness_2024}, Gutzwiller method \cite{hu_link_2024}, scaling theory \cite{chou_scaling_2023,zhou_kondo_2024}, and renormalized perturbation theory \cite{wang_molecular_2024}. 
Within a unified picture, these works provide explanations for experimental phenomena including quantum-dot-like cascade and zero-energy peak seen in STM spectrum, large entropy upon heating, sawtooth compressibility, V-shaped pairing gap, strong coupling feature of the superconductivity, and a link between cascade transitions and correlated Chern insulators.

\begin{table}[ht]
    \centering
    \begin{tabular}{ |c | c | c |}
    \hline 
         &THF model & Ref.~\cite{ledwith_nonlocal_2024} (non-local moments)  \\
         \hline 
         Wave function & $e^{\ii \frac{\pi}4 l} \cdot 
  \frac{ 1 +  \frac{(x+iy)/\lambda}{ \ii l (k_x + ik_y)/(\gamma/v_\star) } }{ \sqrt{ 1 + \frac{\gamma^2/v_\star^2}{\kk^2}  } }
  \cdot \frac{1}{\sqrt{2\pi} \lambda } \sum_{\RR} e^{- \frac{|\rr-\RR|^2}{2\lambda^2}}$ &   $\chi_{l\beta} \cdot \frac{1 + \frac{(x+iy)/\delta}{ i(k_x + ik_y)/s }  }{\sqrt{1 + \frac{2s^2}{\kk^2} }} 
  \cdot \frac1{\sqrt{2\pi}\delta} \sum_{\RR} e^{-\frac{|\rr-\RR|^2}{4\delta^2}} $\\
  \hline 
         Berry curvature &$\Omega_\kk = \frac{ 2\gamma^2/v_\star^2 }{(\gamma^2/v_\star^2+ |\kk|^2)^2}$ &$\Omega_\kk=\frac{4s^2}{(2s^2 +|\kk|^2)^2}$
         \\ 
         \hline 
         Self-energy at $\nu=0$ & $\frac{U_1^2| \kk|^2 }{4 i\omega (| \kk|^2 +\gamma^2/|v_\star|^2) }$ (Hubbard-I approximation) & 
         $\frac{U_1^2|\kk|^2 }{4i\omega (|\kk|^2+2s^2)}$
         \\ 
         \hline 
         Ferromagnetic coupling strength & $\sim -U_1\frac{\gamma^2}{v_\star^2\Lambda_c^2}\log(\frac{|\gamma|}{|v_\star \Lambda_c|})$ & $\sim -U_{\Gamma}s^2\log(s)$ \\
         \hline 
    \end{tabular}
    \caption{To establish the equivalence between our results and those in Ref.~\cite{ledwith_nonlocal_2024}, we let $s = | \gamma/(v_\star\sqrt{2})|$ and $\delta \equiv \lambda$. In addition, Ref.~\cite{ledwith_nonlocal_2024} employs a different convention for the interaction strength. In the table above, we have converted their expressions to match the convention used in our work.
    } \label{tab:comp_table}
\end{table}

Previous work based on the THF model has primarily focused on the parameter region where the Hubbard repulsion ($U_1$) of $f$ electrons exceeds the gap ($\gamma$) between the remote bands and flat bands, {\it i.e.,} $U_1 > \gamma$. 
However, some works\cite{PhysRevB.110.045133} have projected Hamiltonian (such as electron-phonon couplings), onto the flat bands using the wavefunction previously derived in \cite{song_magic-angle_2022}.
In this study, we explore the THF model in the projected $U_1< \gamma$ limit, where the Hubbard repulsion of $f$ electrons is smaller than the gap between the flat bands and remote bands and larger than the bandwidth of flat bands. 
Previous studies of the projected limit have relied on the momentum-space Bistritzer-MacDonald model with projected Coulomb repulsion~\cite{KAN20a,KAN21,VAF20,KAN19,XIE22,KUM21,KWA21,BER21a,LIA21,BER21b,BUL20a,POT21,WU18}, where features of the ``heavy particles'', such as the flat excitation spectrum near the Brillouin zone boundary, are clearly observed \cite{KAN21,BER21a,LIA21,BER21b,BUL20a}.
These features seen in the momentum-space model are naturally explained by the THF model. 
However, a local description of only the flat bands is impossible due to their topological nature. The closest approximation is the nonlocal Wannier function of the topological in Ref.~\cite{ledwith_nonlocal_2024}, which we show shares many similarities with the THF framework, including the wavefunction initially obtained in Ref.~\cite{song_magic-angle_2022} (Eq.~(S123)). In addition, we do not parametrically let the Berry curvature become delta function concentrated, as in that case the gap $\gamma$ would necessarily also vanish, which would invalidate the projection to the flat bands for \emph{any} finite $U_1$.
 
In this work, we explore the correlation effect in the projected limit of the THF model. 
In~\cref{sec:thf_model}, we first review the THF model. In \cref{sec:nonlocal_wannier}, we explore both $k$-space and real-space non-local Wannier functions for topological flat bands within the THF model framework, as previously discussed in Ref.~\cite{song_magic-angle_2022}. 
In~\cref{sec:int_effect}, we explore the correlation effect in the projected limit of the THF.
In~\cref{sec:comp} and in \cref{tab:comp_table}, we provide the comparison between the results obtained from THF model and the results in Ref.~\cite{ledwith_nonlocal_2024}. 
In~\cref{sec:THF-parameter}, we discuss the realistic parameters of the THF model.
Finally, we summarize our results in~\cref{sec:summary}.

\section{The topological heavy fermion model}
\label{sec:thf_model}
The single-particle Hamiltonian of THF model is 
\begin{align} \label{eq:H0}
  \hH_0 =& \sum_{\eta s} \sum_{aa'} \sum_{\kk,\GG} ( H^{(c,\eta)}_{a,a'}(\kk+\GG) - \mu \delta_{aa'} ) c_{\kk+\GG,a,\eta,s}^\dagger c_{\kk+\GG,a'\eta s} - \mu \sum_{\eta s} \sum_{\RR} f_{\RR \alpha\eta s}^\dagger f_{\RR \alpha\eta s}  \nonumber\\ 
& + \frac1{\sqrt{N_M}} \sum_{\eta s \alpha s} \sum_{\RR} \sum_{\kk \GG}
  \pare{ e^{-\frac{\lambda}2 |\kk+\GG|^2 - i\kk\cdot\RR } H_{a\alpha}^{(cf,\eta)}(\kk) 
    c_{\kk+\GG,a,\eta,s}^\dagger f_{\RR \alpha \eta s} + h.c. 
  }\ .
\end{align}
Here $\RR$ sums over all the triangular moir\'e lattice formed by the AA-stacking regions of MATBG, $\kk$ takes value in the moir\'e Brillouin zone, $\GG$ sums over the moir\'e reciprocal lattice, and $N_M$ is the number of moir\'e unit cell. 
$\eta=\pm$ is the index for graphene valleys, $s=\up,\down$ is the spin index, $\alpha=1,2$ is the orbital index for $f$-electrons, and $a=1,2,3,4$ is the orbital index for $c$-electrons. 
The $f$-orbital can be expressed in terms of the continuous basis of the BM model 
\begin{equation} \label{eq:f-basis}
  f_{\RR,\alpha,\eta,s}^\dagger =  \sum_{l\beta} 
  \int d^{2}\rr \  w_{l\beta,\alpha}^{(\eta)}(\rr - \RR) c_{l,\beta,\eta,s}^\dagger(\rr)\ ,
\end{equation}
where $c_{l,\beta,\eta,s}^\dagger(\rr)$ creates a graphene electron at the (coarse-grained and continuous) position $\rr$ in the valley $\eta$, layer $l$, sub-lattice $\beta$, spin $s$. 
Here $\beta=1,2$ corresponds to the A and B sublattices of graphene, respectively; $l=+,-$ corresponds to the top and bottom layer, respectively. 
$w_{l\beta,\alpha}^{(\eta)}(\rr - \RR)$ is an exponentially localized Wannier function with a spread $\lambda$ about $1/2$ of the moir\'e lattice constant. 
The $c$-electrons can also be expressed in terms of the continuous basis as 
\begin{equation} \label{eq:c-basis}
c_{\kk+\GG,a,\eta,s}^\dagger = \frac1{\sqrt{N_M\Omega_M}}
 \sum_{l\beta} \int d^2\rr\ 
 \td{u}^{(\eta)}_{l\beta,a}(\rr;\kk) c_{l\beta \eta s}^\dagger (\rr)\ ,
\end{equation}
where $\Omega_M$ is the area of the moir\'e unit cell. 
Ref.~\cite{song_magic-angle_2022} numerically constructed $\td{u}^{(\eta)}_{l\beta,a}(\rr;\kk)$ around $\kk=0$. 
Ref.~\cite{calugaru_tbg_2023} (in its Appendix F, Eqs. F1-F4)) further provided a Gaussian form for it at momentum away from $\textbf{0}$.

$\mu$ and $\lambda$ in \cref{eq:H0} are the chemical potential and the spread of $f$-orbitals, respectively. 
The hopping matrices $H^{(c,\eta)}(\kk)$ and $H^{(cf,\eta)}(\kk)$  are 
\begin{equation} \label{eq:HcHcf}
  H^{(c,\eta)}(\kk) = 
  \begin{pmatrix}
    0_{2\times} & v_\star (\eta k_x \sigma_0 + ik_y \sigma_z) \\
    v_\star (\eta k_x \sigma_0 - ik_y \sigma_z) & M\sigma_x 
  \end{pmatrix},\qquad 
  H^{(cf,\eta)}(\kk) = 
  \begin{pmatrix}
    \gamma \sigma_0 + v_\star' (\eta k_x \sigma_x + k_y \sigma_y) \\
    0_{2\times2}
  \end{pmatrix}\ .
\end{equation}
The parameters $M$, $v_\star$, $\gamma$, $v_\star'$, $\lambda$ have been extracted numerically and analytically from the BM model in a {\it first-principle spirit} \cite{song_magic-angle_2022,calugaru_tbg_2023}. 
Thus, they are functions of the twist angle $\theta$ and the ratio $w_0/w_1$ between inter-layer hoppings in the AA-stacking regions and AB-stacking regions. 
The two flat bands can hold eight electrons per moir\'e unit cell. 
We use $\nu$ to denote the number of electrons ($\nu>0$) or holes ($\nu<0$) per moir\'e unit cell counted from the charge neutrality point (CNP). 
$\nu=\pm4$ corresponds to the band gaps between the flat bands and remote bands.

We do not give the explicit interaction Hamiltonian $H_I$ here. Interested readers may refer to Ref.~\cite{song_magic-angle_2022} for full details. 
The most important term in $H_I$ is an onsite Hubbard interaction $\hH_{U_1}$
\begin{align}
    \hH_{U_1} = \frac{U_1}{2}\sum_{\RR,\alpha\eta s,\alpha'\eta's'}:f_{\RR,\alpha,\eta,s}^\dag f_{\RR,\alpha,\eta,s}:
    :f_{\RR,\alpha',\eta',s'}^\dag f_{\RR,\alpha',\eta',s'}:
\end{align}
where $U_1$ is the interaction strength and $:f_{\RR,\alpha,\eta,s}^\dag f_{\RR,\alpha,\eta,s}: =f_{\RR,\alpha,\eta,s}^\dag f_{\RR,\alpha,\eta,s}-\frac{1}{2} $. 
Other terms include the density-density interactions between $f$- and $c$-electrons $\hH_W$, density-density interactions between nearest neighbors of $f$-electrons $H_{U_2}$, a Hund's exchange between $f$- and $c$-electrons $\hH_J$, and Coulomb repulsions between $c$ electrons $\hH_V$. 
All the interaction parameters are also determinate functions of the $\theta$ and $w_0/w_1$~\cite{calugaru_tbg_2023}. 
The single-particle and interaction parameters at the magic angle are given in supplementary tables S4 and S6 of Ref.~\cite{song_magic-angle_2022}. The full interacting theory, with terms beyond $U_1$ treated at the mean-field level, is detailed in an extensive DMFT-Iterative Perturbation Theory–Hubbard-I manuscript \cite{DC_DMFT}.




\section{Wavefunction of the topological flat band}
\label{sec:nonlocal_wannier}
Via the THF model~\cite{song_magic-angle_2022}, we could analytically derive the wavefunction for the topological flat bands.
We consider the chiral-flat limit of the THF model by setting $M=0,v_\star'=0$ in \cref{eq:HcHcf}. 
Then, one can obtain exact flat Chern band solutions straightforwardly. 
The creation operator for Bloch state with Chern number $\zeta$ in valley $\eta $ is (Eq.~(S123) of Ref.~\cite{song_magic-angle_2022})
\begin{equation} \label{eq:d-operator}
  d_{\kk,\zeta,\eta,s}^\dagger = \frac1{\sqrt{\mathcal{N}_\kk}} f_{\kk,1,\eta,s}^\dagger 
  + \frac1{\sqrt{\mathcal{N}_\kk}} \sum_{\GG} \frac{\frac{\gamma}{v_\star} e^{-\frac12\lambda^2|\kk+\GG|^2}}{\eta (k_x+G_x) + i(-1)^{\zeta+1}(k_y+G_y)}  c_{\kk+\GG,3,\eta,s}^\dagger \ ,
\end{equation}
where  $\mathcal{N}_\kk = 1 + \sum_{\GG} \frac{(\gamma/v_\star)^2}{|\kk+\GG|^2} e^{-\lambda^2|\kk+\GG|^2}$ is a normalization factor.
To derive the analytical formula of topological flat-band wavefunction, we omit all $|\GG|>0$ components. 
Using \cref{eq:f-basis,eq:c-basis}, we obtain the following Bloch function with Chern number $+1$ in valley $+$
\begin{equation}
\psi_{l\beta;\kk}(\rr) = \sqrt{N_M}\inn{0|c_{l\beta + s}(\rr) d_{\kk,+,+,s}^\dagger|0}
=  \frac1{\sqrt{\mathcal{N}_\kk}} \sum_{\RR} w^{(+)}_{l\beta,1}(\rr-\RR) e^{\ii\kk\cdot\RR}
  + \frac1{\sqrt{\mathcal{N}_\kk}} \frac{\frac{\gamma}{v_\star} e^{-\frac12\lambda^2\kk^2} }{k_x + ik_y} \td{u}_{l\beta,3}^{(+)}(\rr;\kk)\ .
\end{equation}
In the chiral limit, $w^{(+)}_{l\beta,1}(\rr)$ takes the form $w^{(+)}_{l\beta,1}(\rr) = \delta_{\beta 1} \frac1{\sqrt{2\pi} \lambda } e^{i\frac{\pi}4 l - \rr^2/(2\lambda^2)}$ (see Eq.~(S53) of Ref.~\cite{song_magic-angle_2022}). 
Appendix F of Ref.~\cite{calugaru_tbg_2023} also gave the analytical expression for $\td{u}$ 
\begin{align}
\td{u}_{l\beta,3}^{(+)}(\rr;\kk) \approx \delta_{\beta 1} \frac1{\Omega_M} \sum_{\GG}  e^{-\ii\frac{\pi}4 l  - \frac12 \qq^2 \lambda^{\prime 2} + \ii (\kk + \GG)\cdot\rr } (\ii G_x - G_y ) \sqrt{2\pi \lambda^{\prime 4} }
    \ , \qquad 
(\text{for}\; |\kk| a_M\ll 1)
\end{align}
where $a_M$ is the moir\'e lattice constant. 
Here $\lambda'$ characterizes the spread of the $c$-electrons. 
It is usually larger than $\lambda$ as $c$-electron is less localized. 
Inserting the identity $ \frac1{\Omega_M}\sum_{\GG} \delta(\qq - \GG) = \frac1{(2\pi)^2}\sum_{\RR} e^{i \qq\cdot\RR}$, we have 
\begin{align}
\td{u}_{l\beta,3}^{(+)}(\rr;\kk) =& \delta_{\beta 1} \sum_{\RR} \int \frac{d^2\qq}{(2\pi)^2} e^{-\ii\frac{\pi}4 l  - \frac12 \qq^2 \lambda^{\prime 2} + \ii (\kk+\qq)\cdot\rr - \qq\cdot\RR } (\ii q_x - q_y ) \sqrt{2\pi \lambda^{\prime 4} } \nonumber\\
=& \delta_{\beta 1} \frac{1}{\sqrt{2\pi} \lambda^{\prime 2}}  
  \sum_{\RR} e^{-\ii \frac{\pi}4 l + \ii \kk\cdot(\rr-\RR) - \frac{(\rr-\RR)^2}{2\lambda^{\prime2}} }
  (x - R_x + i(y-R_y) )\ , \qquad (|\kk|a_M\ll1)\ . 
\end{align}
To simplify the expression, we take take $\lambda'=\lambda$ which leads to a $\kk$-independent Gaussian factor $e^{-\frac{|\rr-\RR|^2}{2\delta^2}}$. 
As $\lambda$ is smaller than $a_M$, there are $\kk^2\lambda^2\ll 1$ for $\kk$ in first Brillouin zone and $\lambda^2\ll \RR^2$ for nonzero $\RR$. 
Then we obtain the wave function in first Brillouin and first unit cell as 
\begin{equation} \label{eq:psi}
\psi_{l\beta;\kk}(\rr) = \delta_{\beta 1} e^{\ii \frac{\pi}4 l} \cdot 
  \frac{ 1 +  \frac{(x+iy)/\lambda}{ \ii l (k_x + ik_y)/(\gamma/v_\star) } }{ \sqrt{ 1 + \frac{\gamma^2/v_\star^2}{\kk^2}  } }
  \cdot \frac{1}{\sqrt{2\pi} \lambda } \sum_{\RR} e^{- \frac{|\rr-\RR|^2}{2\lambda^2}}\ .
\end{equation}
which leads to a concentrated charge distribution.

In addition, we calculate the Berry curvature of the flat band. Starting from the ansatz given in~\cref{eq:d-operator}, we make two additional approximations:
(1) we omit all $|\GG|>0$ component; (2) we drop the exponential factor by setting $\lambda=0$. 
This results in the following simplified ansatz:
\begin{align}
\label{eq:d_op_simp}
     d_{\kk,1,+,s}^\dagger \approx 
     \frac1{\sqrt{1+ \frac{\gamma^2}{|v_\star \kk|^2 }}} f_{\kk,1,+,s}^\dagger  
  + \frac1{\sqrt{1+ \frac{\gamma^2}{|v_\star \kk|^2 }}} \frac{\frac{\gamma}{v_\star}}{k_x + ik_y }  c_{\kk+\GG,3,+,s}^\dagger \ ,
\end{align}
Using~\cref{eq:d_op_simp}, we could obtain the following expression for the concentrated Berry curvature of the Chern $+1$ band in valley $+$
\begin{align}
\label{eq:berry_curv_d_op_simp}
    \Omega_{\kk} = \frac{ 2\gamma^2v_\star^2 }{(\gamma^2+v_\star^2 |\kk|^2)^2}\ .
\end{align}
From~\cref{eq:psi,eq:berry_curv_d_op_simp}, we observe that our wavefunction of topological flat bands has two main features:
concentrated charge in real space and concentrated Berry curvature in momentum space.

A similar wavefunction of the topological flat band has also been derived in Ref.~\cite{ledwith_nonlocal_2024}. 
The wavefunction derived in Ref.~\cite{ledwith_nonlocal_2024} for Chern number $\zeta=1$ valley $\eta=+$ takes the form of 
\begin{equation} \label{eq:NW-ansatz}
  \psi^{\rm NW}_{l \beta ; \kk}(\rr) = \chi_{l\beta} \cdot \frac{1 + \frac{(x+iy)/\delta}{ i(k_x + ik_y)/s }  }{\sqrt{1 + \frac{2s^2}{\kk^2} }} 
  \cdot \frac1{\sqrt{2\pi}\delta} \sum_{\RR} e^{-\frac{|\rr-\RR|^2}{4\delta^2}} \ , 
\end{equation} 
where $\chi_{l\beta}$ is a spinor in valley and sub-lattice space, $\rr\!=\!(x,y)$ is restricted in the first moir\'e unit cell, and $\kk$ is restricted in the first moir\'e Brillouin zone.  
Here $\delta$ controls the spread of the concentrated charge in AA stacking regions, and $s$ controls the spread of Berry curvature in momentum space. 
When $\rr$ ($\kk$) exceeds the first unit cell (Brillouin zone), periodic conditions $\psi_{l\beta;\kk+\GG}^{\rm NW}(\rr)=\psi_{l\beta;\kk}^{\rm NW}(\rr)$, $\psi_{l\beta;\kk+\GG}^{\rm NW}(\rr+\RR)=e^{\ii\kk\cdot\RR}\psi^{\rm NW}_{l\beta;\kk}(\rr)$ are enforced. 
A subtlety of \cref{eq:NW-ansatz} is that $|\psi^{\rm NW}_{l\beta;\kk}(\rr)|^2$ exhibits artificial discontinuities in both momentum and real spaces. 
In Ref.~\cite{ledwith_nonlocal_2024} $\delta$ and $s$ are treated as fitting parameters to minimize the error in the intra-Chern-band form factors. The inter-Chern-band form factors are omitted. 
This automatically leads to an artificial chiral symmetry as well as a U(4)$\times$U(4) symmetry in the projected interaction, as explained in Ref.~\cite{bernevig_twisted_2021}, even though the parameters $\delta,s$ are fitted using realistic parameters without chiral symmetry ($w_0\neq 0$). 
We termed such projected interaction as in a {\it quasi-chiral} limit which has also been discussed in Ref.~\cite{song_magic-angle_2022}. 

Comparing \cref{eq:psi} to \cref{eq:NW-ansatz}, we identify the correspondence between THF parameters and NW ansatz parameters:
\begin{equation}
  s\equiv |\frac{\gamma}{\sqrt{2} v_\star}|,\qquad 
  \delta \equiv \lambda \ .
\end{equation} 
Our form in~\cref{eq:psi} obtained in the chiral-flat limit differs from the NW ansatz (\cref{eq:NW-ansatz}) by a $\ii l$ phase factor. 
Through the THF model, we could also go beyond the chiral-flat limit by introducing non-zero $M$ and $v_\star^\prime$ and obtain the flat-band wave function either numerically or perturbatively. Additionally, by including $c$-electrons for all the $\GG$ shell in~\cref{eq:d-operator}, our derived expressions~\cite{song_magic-angle_2022} for the wavefunction of flat bands do not exhibit any discontinuities. \cref{eq:psi} (as also derived in Ref.~\cite{ledwith_nonlocal_2024}) can be understood as a special case of it.

\section{Correlation effects in the projected limit of the topological heavy-fermion model}
\label{sec:int_effect}

In this section, we study the correlation effects of the heavy-fermion model in the projected limit 
\begin{align}
\label{eq:weak_hyb_limit}
  |v_\star \Lambda_c| \gg   |\gamma| \gg  U_1 
\end{align}
and demonstrate the local moment formation in this limit of the THF model. 
Here, $\Lambda_c$ is the momentum cutoff of the $c$ electron, which is on the order of $\sim \frac{1}{2}|\mathbf{b}_{M,1}|$. Thus, $  |v_\star \Lambda_c| $ can be interpreted as the bandwidth of $c$ electrons. In the projected limit, the gap between the flat band and remote band, $\gamma$, is larger than the Hubbard interaction $U_1$.
Only the flat bands remain active and experience strong interaction effects. Moreover, since $\gamma/ |v_\star \Lambda_c| \ll  1 $, we only expect a non-zero Berry curvature in a small (but non-vanishing) region near $\Gamma$ point, where $|\kk| < \frac{|v_\star|}{\gamma}$ (see~\cref{eq:berry_curv_d_op_simp}). 
In other words, the band remains trivial over most of the moir\'e Brillouin zone.
In addition, the orbital weight of $f$ electrons in the flat band is (from~\cref{eq:d_op_simp})
\begin{align}
   \frac{1}{N_M} \sum_{\kk }\frac{1}{1+ \frac{\gamma^2}{|v_\star \kk|^2}  } \approx \frac{1}{\Omega_{M}} 
   \int_{|\kk|<\Lambda_c}\frac{1}{1+ \frac{\gamma^2}{|v_\star \kk|^2}  } 
   \label{eq:approx_sum_to_int}
\end{align}
with $\Omega_{M}$ the size of the moir\'e Brillouin zone. In practice, we can set $\Omega_{M} \approx \pi \Lambda_c^2$. 
In other words, we approximate the original hexagonal Brillouin zone with a circular one. This leads to the following orbital weights for the $f$ electrons:
\begin{align}
     \frac{1}{N_M} \sum_{\kk }\frac{1}{1+ \frac{\gamma^2}{|v_\star \kk|^2}  } \approx 1 + \frac{\gamma^2}{| v_\star \Lambda_c|^2}
     \log\frac{1}{1+ \frac{|v_\star \Lambda_c|^2}{\gamma^2}}
     \label{eq:orb_weight_f_in_flat_band}
\end{align}
For $|\gamma|/|v_\star \Lambda_c| \ll 1 $, the orbital weight of $f$ electrons is $\sim 1 -2\bigg[|\gamma|/|v_\star \Lambda_c|\bigg]^2\log(|\gamma /|v_\star \Lambda_c|)  $, indicating that the flat bands are mostly formed by $f$ electrons in this limit. Since, in the projected limit, only the flat bands are the relevant low-energy degrees of freedom and these bands are primarily composed of $f$ electrons, we expect the correlation effects to be well described by the heavy-fermion $f$-electron basis and largely governed by the Hubbard interactions between $f$ electrons ($\hH_{U_1}$).

In addition, the local moment formation can be understood by viewing each $f$-site as an Anderson impurity coupled to an effective bath describing its environment. 
Ref.~\cite{zhou_kondo_2024} (Eq.~(B25)) derived the hybridization function of the effective Anderson impurity in the flat-band limit ($M=0$)
\begin{equation}
    \Delta(\omega) = \frac{\sqrt3}{8} \cdot \frac{a_M^2 \gamma^2}{ v_\star^2} |\omega|\ ,
\end{equation}
where $a_M\sim \Lambda_c^{-1}$ is the moir\'e lattice constant. 
As revealed in many calculations, the bandwidth of the renormalized spectrum is driven by $U_1$ (but can be renormalized down, to for example $0.6 U_1$ in DMFT results \cite{rai_dynamical_2023}).
Thus, the typical hybridization within this energy scale, which characterizes the inversion lifetime of $f$-electrons, is given by $\Delta_0(U_1)$. 
A local moment can be stabilized as long as 
\begin{equation}
    U_1 \gg \Delta_0 (U_1)\qquad \Leftrightarrow \qquad 
    \frac{a_M\gamma}{v_\star} \ll 1 
    \quad \text{or} \quad 
    \frac{\gamma}{v_\star \Lambda_c} \ll 1\ .  
\end{equation}
The Kondo temperature $T_{\rm K} \propto \exp(-\frac{\pi U_1}{32\Delta})$ is exponentially suppressed in this limit. Since $    \frac{\gamma}{v_\star \Lambda_c} \ll 1 $ implies $\gamma < 100 meV$,  local moment formation happens in TBG.

In this section, we discuss the correlation effects in the projected limit (\cref{eq:weak_hyb_limit}). 
We derive the self-energy of the flat bands, and also the effective coupling between flavor moments formed by the flat bands. 
To simplify the analysis, we will take the chiral flat limit with $v_\star^\prime =0, M=0$. 
Since the flat bands are primarily composed of $f$ electrons, we expect the correlation effects to be driven by the Hubbard interaction of the $f$ electrons.
Therefore, we only treat $H_W,H_V$ at the Hartree-Fock level, where they contribute only single-particle terms to the Hamiltonian.
We will also drop the $H_J$ which is generally much weaker than the other interactions.


\subsection{Self-energy and Hubbard-I approximation}
To obtain the self-energy, we first integrate out the $c$ electrons, leading to the following effective action of $f$ electrons
\begin{align} \label{eq:Sf}
    S_f =& \frac{1}{\beta} \sum_{\RR,\alpha\eta s,\omega}f_{\RR,\alpha \eta s}^\dag(i\omega) 
    \big(i\omega - \Sigma_0(\RR-\RR',\ii\omega) + \epsilon_f \big)
    f_{\RR,\alpha \eta s}(i \omega)
    \nonumber   \\
& + \frac{U_1}{2}\int_0^{\beta} d\tau \sum_{\RR,\alpha\eta s,\alpha'\eta's'}       
    \pare{f_{\RR,\alpha\eta s}^\dag (\tau) f_{\RR,\alpha\eta s} (\tau) -\frac12 }
     \pare{f_{\RR,\alpha'\eta' s'}^\dag (\tau) f_{\RR,\alpha'\eta' s'} (\tau) - \frac12} 
\end{align}
Here, the $\Sigma_0$ term denotes the contribution from integrating out $c$ electrons, $\epsilon_f $ is the chemical potential term of $f$ electrons which includes the Hartree contributions from $H_{U_2},H_{W}$~\cite{hu_kondo_2023}, $\tau$ is imaginary time, $\omega$ is Matsubara frequency, and $\beta$ is the inverse temperature. 
Explicitly, $\Sigma_0$ is given by
\begin{align}
\Sigma_0(\RR-\RR',i\omega)
=& \frac{1}{\Omega_{M}} \int_{|\kk|<\Lambda_c} \frac{d^2\kk}{(2\pi)^2}
  \bigg[ \frac{\gamma^2/2}{\ii\omega - |v_\star \kk|-\epsilon_c } 
    +\frac{\gamma^2/2}{\ii\omega + |v_\star \kk|-\epsilon_c } \bigg]e^{i \kk\cdot(\RR-\RR')} \nonumber\\
=& \frac{\gamma^2}{v_\star^2\Lambda^2} \frac{2\pi}{\Omega_{M}} \int_{\epsilon < |v_\star \Lambda_c|} d\epsilon 
    \brak{  \frac{\epsilon/2}{\ii\omega + \epsilon_c - \epsilon } 
    +\frac{\epsilon/2}{\ii\omega + \epsilon_c + \epsilon }    } J_0\pare{ \frac{\epsilon \cdot |\RR-\RR'|}{|v_\star|} }
    \label{eq:c_self_energy_contribution}
\end{align}
where $\epsilon_c$ describes the Hartree contributions to the $c$ electrons from the $\hat{H}_W, \hat{H}_V$ \cite{DC_DMFT,rai_dynamical_2023,hu_kondo_2023}. From~\cref{eq:c_self_energy_contribution}, we observe the contribution from $c$ electrons approaches zero as $\frac{\gamma^2}{v_\star^2\Lambda_c^2}\rightarrow 0$. However, it is worth mentioning that the exact limit $\frac{\gamma^2}{v_\star^2\Lambda_c^2}= 0$ also indicates the gap vanishes ($\gamma\rightarrow 0$), given that $v_\star , \Lambda_c$ are finite. 

\begin{figure}
    \centering
    \includegraphics[width=1.0\linewidth]{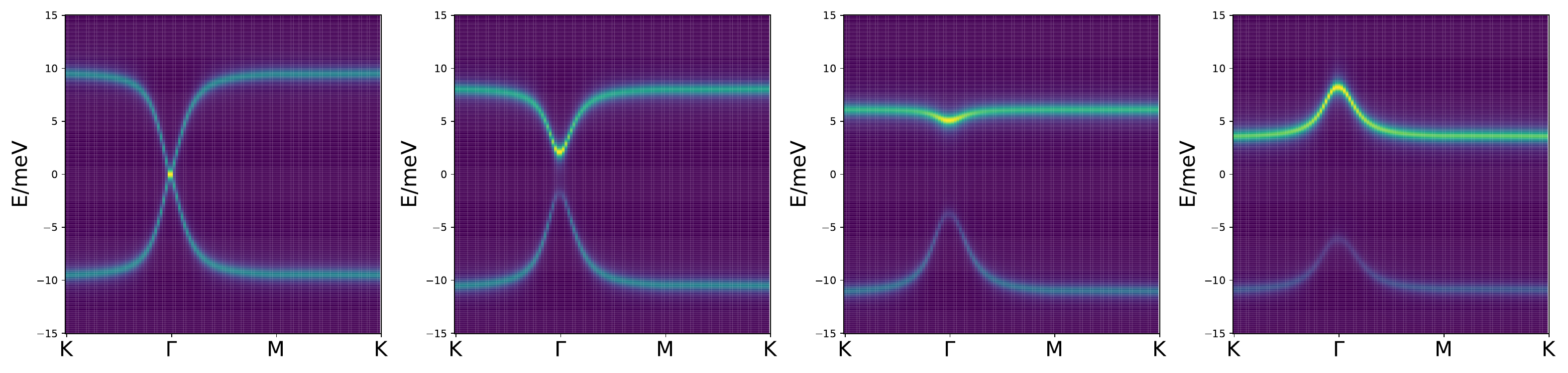}
    \caption{Spectrum obtained from Hubbard-I approximation at integer fillings $\nu=0,-1,-2,-3$ (from left to right). To realize the limit of $|v_\star \Lambda_c|\gg \gamma \gg U_1$, we have taken the parameter values given in Ref.~\cite{song_magic-angle_2022}, and then rescaled all the interaction strength by a factor of $1/3$. }
    \label{fig:hubbard_I_spec}
\end{figure}

We first derive the self-energy of $f$ electrons in the limit of $|\gamma|/|v_\star \Lambda_c| \rightarrow 0$. In this limit, the effective action of $f$ electrons is
\begin{align}
\label{eq:eff_action_atomic_limit}
 S_{at} =    \frac{1}{\beta} \sum_{\RR,\alpha\eta s,i\omega}f_{\RR,\alpha \eta s}^\dag(i\omega) (i\omega - \epsilon_f) f_{\RR,\alpha \eta s}(i\omega)
+\frac{U_1}{2}\int_\tau \sum_{\RR,\alpha\eta s,\alpha'\eta's'} :f_{\RR,\alpha\eta s}^\dag (\tau):f_{\RR,\alpha\eta s} (\tau):
     :f_{\RR,\alpha'\eta' s'}^\dag (\tau):f_{\RR,\alpha'\eta' s'} (\tau):
\end{align}
This describes an atomic problem where $f$ electrons of different sites are decoupled. 
We use $\nu_f$ to characterize the filling of the $f$ electrons. The atomic action $S_{at}$ can be solved exactly which yields the following self-energy in the low-energy limit for the $f$ electrons at the integer filling ($1/\beta \ll U_1$)~\cite{DC_DMFT}
\begin{align}
    \Sigma_f(i\omega) = \frac{1}{64}\frac{ U_1^2 (4-\nu_f)(\nu_f+4)}{ i \omega - U_1\nu_f/8}
\end{align}
Moreover, in the projected limit, the difference between the filling of $f$ electrons, $\nu_f$, and the total filling of the systems, $\nu$, is of the order $\gamma^2/|v_\star \Lambda_c|^2\log(|\gamma|/|v_\star \Lambda_c|)$, as estimated from the orbital weights of $f$ electrons of the flat bands (\cref{eq:orb_weight_f_in_flat_band}). Thus, as $|\gamma|/|v_\star \Lambda_c| \rightarrow 0$, the self-energy of $f$ electrons can be approximately written as 
\begin{align}
\label{eq:self-energy-fele}
     \Sigma_f(i\omega) \approx  \frac{1}{64}\frac{ U_1^2 (4-\nu)(\nu+4)}{ i \omega - U_1\nu/8}
\end{align}
Using~\cref{eq:self-energy-fele}, we can now investigate the single-particle excitation by evaluating the interacting Green's functions. 
We define the full Green's function as 
\begin{align}
    G^{\alpha \eta s}(\tau,\kk) = 
    \begin{bmatrix}
        -\langle T_\tau f_{\kk,\alpha \eta s}(\tau) f_{\kk,\alpha \eta s}^\dag \rangle 
        &  -\langle T_\tau f_{\kk,\alpha \eta s}(\tau) c_{\kk,\alpha \eta s}^\dag \rangle 
        &  -\langle T_\tau f_{\kk,\alpha \eta s}(\tau) c_{\kk,\alpha+2 \eta s}^\dag \rangle \\ 
       -\langle T_\tau c_{\kk,\alpha \eta s}(\tau) f_{\kk,\alpha \eta s}^\dag \rangle  
       &-\langle T_\tau c_{\kk,\alpha \eta s}(\tau) c_{\kk,\alpha \eta s}^\dag \rangle  
        &-\langle T_\tau c_{\kk,\alpha \eta s}(\tau) c_{\kk,\alpha+2 \eta s}^\dag \rangle  \\ 
    -\langle T_\tau c_{\kk,\alpha+2 \eta s}(\tau) f_{\kk,\alpha \eta s}^\dag \rangle  
    &-\langle T_\tau c_{\kk,\alpha+2 \eta s}(\tau) c_{\kk,\alpha \eta s}^\dag \rangle  
    &-\langle T_\tau c_{\kk,\alpha+2 \eta s}(\tau) c_{\kk,\alpha+2 \eta s}^\dag \rangle  
    \end{bmatrix}
\end{align}
Using the self-energy in~\cref{eq:self-energy-fele}, we obtain
\begin{align}
    [G^{\alpha \eta s}(i\omega,\kk)]^{-1} = 
    \begin{bmatrix}
        i\omega -\epsilon_f& -\gamma & 0  \\
       -\gamma  & i\omega -\epsilon_{c} & - v_\star (\eta k_x + i(-1)^{\alpha+1} k_y)\\
       0 & -  v_\star (\eta k_x -i(-1)^{\alpha+1} k_y) & i\omega -\epsilon_{c} 
    \end{bmatrix} - 
    \begin{bmatrix}
      \frac{1}{64}\frac{ U_1^2 (4-\nu)(\nu+4)}{ i \omega - U_1\nu/8}& 0 & 0 \\ 
        0 & 0 & 0 \\ 
        0 & 0 & 0 
    \end{bmatrix}
    \label{eq:hubbard_I_green_fun}
\end{align} 
The Green's function in the above equation with atomic self-energy is equivalent to the Green's function obtained from Hubbard-I approximation~\cite{hubbard1963electron}. 
We can determine the Hartree contributions $\epsilon_f,\epsilon_{c}$ self-consistently at each filling, and then numerically obtain the spectral functions $\rho(\epsilon) =\frac{1}{\pi} \text{Tr}[G^{\alpha\eta s}(\omega-i0^+)]$ (\cref{fig:hubbard_I_spec}). We show the resulting spectral functions at $\nu=0,-1,-2,-3$ in~\cref{fig:hubbard_I_spec}, which is consistent with the result derived in Ref.~\cite{ledwith_nonlocal_2024} via the momentum-space model.

We now note that, via the THF model, we can also evaluate the excitation spectrum \emph{analytically}. 
Since we are in the limit of $\gamma \gg U_1$, we can project the Green's function to the flat band. The wavefunctions of the flat bands are (see also~\cref{eq:d_op_simp})
\begin{align}
\label{eq:d_op_simp_2}
    v^{\alpha \eta s}_\kk = \frac{1}{\sqrt{1+\gamma^2/|v_\star \bf{k}|^2}}
    \begin{bmatrix}
         1  & 0
         & \frac{\gamma}{v_\star (\eta k_x +i (-1)^{\alpha+1}k_y)}
    \end{bmatrix}^T
\end{align}
It is worth mentioning that, away from CNP, non-zero $\epsilon_f$ and  $\epsilon_c$ appear, which could modify the wave function of flat bands in~\cref{eq:d_op_simp_2}. However, since $\epsilon_f,\epsilon_c$ are both proportional to the interaction strength and are much smaller than $\gamma$ under our assumptions. In ~\cite{hubbard1963electron} we give the expressions for the Hubbard-I for all range of  parameters $\gamma, U_1$. We expect the flat-band wavefunction can still be approximately written as~\cref{eq:d_op_simp_2}. 
To get a very simple analytical formula, we  approximately set $\epsilon_f=\epsilon_c=0$ and project the Green's function to the flat bands. This gives the following interacting Green's function of flat bands
\begin{align}
    [G_{\rm flat}^{\alpha \eta s}(i\omega,\kk)]^{-1} = v_\kk^{\alpha\eta s,\dag} \cdot  [G^{\alpha\eta s}(\omega,\kk)]^{-1}\cdot  v_\kk^{\alpha\eta s} = 
    i\omega - \frac{U_1^2|v_\star \kk|^2 (4-\nu)(4+\nu)}{64(i\omega-U_1\nu/8) (|v_\star \kk|^2 +\gamma^2) }
\end{align}
This also indicates the self-energy of the flat-bands can be written as 
\begin{align}
   \Sigma_{flat}(i\omega)= \frac{U_1^2|v_\star \kk|^2 (4-\nu)(4+\nu)}{64(i\omega-U_1\nu/8) (|v_\star \kk|^2 +\gamma^2)  }
\end{align}

We could focus on $\nu=0$, the self-energy then becomes 
\begin{align}
     \Sigma_{flat}(i\omega)= \frac{U_1^2|v_\star \kk|^2 }{4 i\omega (|v_\star \kk|^2 +\gamma^2) }
\end{align}
The single-particle excitation of the system can be obtained by finding the poles of Green's function ($[G^{\alpha\eta s}_{flat}(i\omega \rightarrow E_\kk, \kk) ]^{-1}= 0 $) which gives (at $\nu=0$)
\begin{align}
    E_{\kk,1/2} = \pm \frac{1}{2} \sqrt{ \frac{ U_1^2 |v_\star \kk|^2}{|v_\star \kk|^2 + \gamma^2 }}  
\end{align}
Away from $\Gamma$ point with $|v_\star \kk| \gg \gamma$, the single-particle excitation shows Mott-like behavior and are gapped with $E_{\kk,1/2} \approx \pm \frac{1}{2}U_1 $. Near the $\Gamma$ point with $|v_\star \kk| \ll \gamma$, we obtain a Dirac dispersion with
\begin{align}
\label{eq:dispersion_of_dirac_nu_0}
    E_{\kk,1/2} \approx \pm \frac{U_1}{2|\gamma| }|v_\star \kk| 
\end{align} 
At $\nu \ne 0$, due to the absence of particle-hole symmetry, the Dirac nodes will be gapped (\cref{fig:hubbard_I_spec}).

Finally, the effective action $S_{at}$ (\cref{eq:eff_action_atomic_limit}) describes an atomic problem with Hubbard interactions. 
The entropy of this atomic problem can be solved exactly. At filling $\nu$, the number of $f$ electrons are $\nu_f +4 \approx \nu+4$. These $f$ electrons can occupy $8(=2\times 2 \times 2)$-degenerate $f$ orbitals. The ground states thus have a  degeneracy of $\frac{8!}{(\nu+4)!(4-\nu)!}$. This gives the following entropy of the entire (unstrained) system
\begin{align}
    S = k_BN_M \log( \frac{8!}{(\nu+4)!(4-\nu)!} )
\end{align} The entropy of the THF with strain is solved in \cite{entropy_stain}.

Finally, we comment that we expect the Hubbard-I approximation to provide a qualitative good description of the single-particle dispersion, for both $U_1>|\gamma|$ and $U_1<|\gamma|$, as long as the system does not develop the Kondo effect or long-range ordering. In the low-temperature limit, long-range orders and the Kondo effect become important and can  lead to a qualitative change of the self-energy behaviors as shown in the full DMFT solution \cite{rai_dynamical_2023}. 

\subsubsection{Single-particle spectrum in the Hubbard-I approximation across all interaction regimes}
In this section, we discuss the single-particle spectrum in the Hubbard-I approximation in detail. 
We first utilize the idea introduced in \cref{app:sec:hubbar_I:mapp_green_fun} and map the interacting Green's function to an effective non-interacting Green's function. We note that the self-energy in \cref{eq:self-energy-fele} can be written as 
\begin{equation}
{\Sigma}_{f} \left( i \omega \right)  = \int_{-\infty}^{\infty} 
\frac{\rho^{{\Sigma}_{f}}(\epsilon)}{i\omega-\epsilon} d\epsilon 
,\quad \rho^{{\Sigma}_{f}}(\epsilon)= \frac{U_1^2(4-\nu_f)(\nu_f+4)}{64}\delta\left(\epsilon - U_1\nu_f/8\right)
\end{equation}
Therefore, we can introduce the following non-interacting systems and the auxiliary fermion operators $a$ (similarly as ~\cref{eq:app:aux_ham}) 
\begin{align}
    \hat{H}^{aux} = &\sum_{\vk,\alpha \eta s} [\psi_{\vk,\alpha \eta s}^\dag]_i h_{ij}^{\alpha\eta s}(\vk) [\psi_{\vk,\alpha\eta s}]_j\nonumber\\ 
    & +\sum_{\vk, \alpha \eta s } U_1\nu_f/8 a_{\vk, \alpha \eta s}^\dag a_{\vk, \alpha \eta s}
    + \sum_{\vk, \alpha\eta s} \frac{U_1\sqrt{(4-\nu_f)(\nu_f+4)}}{8}\left( a_{\vk,\alpha\eta s}^\dag {f}_{\vk, \alpha \eta s}+\text{h.c.}\right).
    \label{eq:app:hubbard_i_aux_ham}
\end{align}
where 
\begin{align}
    &\psi_{\kk,\alpha\eta s}= \begin{bmatrix}
        f_{\kk,\alpha\eta s} & c_{\kk,\alpha\eta s}& c_{\kk,(\alpha+2) \eta s}
    \end{bmatrix} \nonumber\\ 
    & h^{\alpha\eta s}(\vk) = 
    \begin{bmatrix}
       \epsilon_f& \gamma & 0  \\
       \gamma  &\epsilon_{c} & v_\star (\eta k_x + i(-1)^{\alpha+1} k_y)\\
       0 &   v_\star (\eta k_x -i(-1)^{\alpha+1} k_y) & \epsilon_{c} 
    \end{bmatrix}
\end{align}
\cref{eq:app:hubbard_i_aux_ham} gives exactly the same Green's function (for $f,c$ electrons) and the same single-particle spectrum as the Hubbard-I Green's function ~\cref{eq:hubbard_I_green_fun}.  
Therefore, it is sufficient to analyze the single-particle spectrum of $\hat{H}^{aux} $ which can be described by the following single-particle Hamiltonian 
\begin{align}
\label{eq:aux_ham_flat_chiral_limit}
    h^{aux,\alpha\eta s}(\kk) =
  \begin{bmatrix}
       \epsilon_f& \gamma & 0  & \frac{U_1\sqrt{(4-\nu_f)(4+\nu_f)}}{8}  \\
       \gamma  &\epsilon_{c} & v_\star (\eta k_x + i(-1)^{\alpha+1} k_y) &0 \\
       0 &   v_\star (\eta k_x -i(-1)^{\alpha+1} k_y) & \epsilon_{c} &0   \\ 
        \frac{U_1\sqrt{(4-\nu_f)(4+\nu_f)}}{8} & 0 & 0 & U_1\nu_f/8
    \end{bmatrix}
\end{align} This is true in all regimes of $U_1$.
Our mapping to the effective system with auxiliary fermions reproduces the original Green's function, which indicates the mapping preserves the symmetries. Since $f_{\kk,\alpha\eta s}$ and $a_{\kk,\alpha\eta s}$ are hybridized with a momentum independent hybridization term, which indicates $a_{\kk,\alpha\eta s} $ and $f_{\kk,\alpha \eta s}$ shares the same symmetry properties. 

We first investigate the spectrum near $\Gamma$ point. The spectrum could be obtained by diagonalizing $H^{aux,\alpha\eta s}(\kk=0)$. In the limit of $U_1 \ll \gamma$, the $f_{\alpha\eta s}, c_{\kk,\alpha\eta s}$ are high energy degrees of freedom and are gapped by large $|\gamma|$. The low-energy bands are described by $c_{\kk,(\alpha+2)\eta s}, a_{\kk,\alpha\eta s}$. We can then treat the effect of $f_{\alpha\eta s},c_{\kk,\alpha\eta s}$ perturbatively by integrating them out. This then leads to the following effective Hamilton 
\begin{align}
&H^{eff} = \sum_{\kk,\alpha\eta s}
\begin{bmatrix}
    c_{\kk,(\alpha+2)\eta s}^\dag & a_{\kk,\alpha\eta s}^\dag
\end{bmatrix} 
h^{eff,\alpha \eta s}(\kk)
\begin{bmatrix}
c_{\kk,(\alpha+2)\eta s}\\ a_{\kk,\alpha\eta s}
\end{bmatrix}\nonumber\\ 
&
h^{eff,\alpha\eta s}(\kk) = 
\begin{bmatrix}
    \epsilon_c & \\ 
    &\frac{U_1 \nu_f}{8} \\ 
\end{bmatrix}  + \frac{-U_1 v_\star \sqrt{(4-\nu_f)(4+\nu_f)}}{8\gamma}
\begin{bmatrix}
   0 &\eta k_x -i (-1)^{\alpha+1}k_y \\ 
\eta k_x +i (-1)^{\alpha+1}k_y& 0
\end{bmatrix}
\end{align}
The corresponding spectrum is 
\begin{align}
    \frac{8\epsilon_c +U_1\nu_f}{16} 
    \pm \sqrt{ \bigg[\frac{8\epsilon_c -U_1\nu_f}{8} 
    \bigg]^2 
    + \frac{U_1^2 v_\star^2(4-\nu_f)(4+\nu_f)}{16^2\gamma^2}|\kk|^2  
    }
\end{align}
We can observe that at charge neutrality, the system is gapless. However, away from charge neutrality, due to the effect of particle-hole symmetry breaking, the system is gapped. 

We could further discuss the single-particle spectrum away from the flat band limit by including a finite $M$ term. 
Our effective Hamiltonian now becomes
\begin{align}
    &H^{eff,M} = \sum_{\kk,\alpha\eta s}
\begin{bmatrix}
    c_{\kk,3\eta s}^\dag & a_{\kk,1\eta s}^\dag
    &
    c_{\kk,4\eta s}^\dag & a_{\kk,2\eta s}^\dag
\end{bmatrix} 
h^{eff,\alpha \eta s}(\kk)
\begin{bmatrix}
 c_{\kk,3\eta s} \\ a_{\kk,1\eta s}
    \\
    c_{\kk,4\eta s} \\ a_{\kk,2\eta s}
\end{bmatrix}\nonumber\\ 
&
h^{eff,\alpha\eta s}(\kk) = 
\begin{bmatrix}
    \epsilon_c & 0 & M  &0 \\ 
 0   &\frac{U_1 \nu_f}{8} &0 & 0\\  
     M &0& \epsilon_c & 0 \\
    0 & 0 & 0 & \frac{U_1 \nu_f}{8}
\end{bmatrix}  + \frac{-U_1 v_\star \sqrt{(4-\nu_f)(4+\nu_f)}}{8\gamma}
\begin{bmatrix}
   0 &\eta k_x -i k_y  & 0 & 0\\ 
\eta k_x +i k_y& 0 & 0 & 0 \\
0 & 0 & 0 & \eta k_x +i k_y\\
0 & 0 & \eta k_x-i k_y &0
\end{bmatrix}
\end{align}
The eigenvalues can be obtained exactly which are
\begin{align}
     &\frac{\epsilon_c + U_1\nu_f/8}{2} +\frac{M}{2} 
    \pm \sqrt{ \tilde{V}^2|\kk|^2 + \frac{1}{4}\bigg( \epsilon_c -U_1\nu_f/8 + M\bigg)^2 } \nonumber\\ 
    & \frac{\epsilon_c + U_1\nu_f/8}{2} - \frac{M}{2} 
    \pm \sqrt{ \tilde{V}^2|\kk|^2 + \frac{1}{4}\bigg( \epsilon_c -U_1\nu_f/8 - M\bigg)^2 } \nonumber\\ 
\end{align}
where 
\begin{align}
    \tilde{V} = \frac{U_1\sqrt{(4-\nu_f)(4+\nu_4)}}{8\gamma}v_\star \,
\end{align}
We now discuss the evolution of Dirac nodes. At charge neutrality with $\nu_f =0,\epsilon_c=0$, the spectrum becomes
\begin{align}
    \pm \bigg( \frac{M}{2} \pm \sqrt{ \tilde{V}^2 |\kk|^2 +M^2/4}\bigg) 
\end{align}
which leads to a quadratic Dirac node. 

Away from charge neutrality, the spectrum at $\kk=0$ is
\begin{align}
\epsilon_c +M ,\quad \epsilon_c -M ,\quad \frac{U_1\nu_f}{8},\quad \frac{U_1\nu_f}{8}
\end{align}
The two-fold degenerate states give a quadratic Dirac node. However, the energy of the Dirac node is no longer at $E=0$ due to the particle-hole symmetry breaking at $\nu\ne 0 $.  Therefore, depending on the parameters of the model, we could have either a gap state or a gapless state. In practice, if 
\begin{align}
\label{eq:gap_condition}
    &\frac{U_1}{8}\nu_f > \epsilon_c+ M > \epsilon_c-M \nonumber\\ 
\text{or}\quad &    \frac{U_1}{8}\nu_f< \epsilon_c+ M > \epsilon_c-M \, ,
\end{align}
then the quadratic Dirac node is formed by the top two bands or the bottom two bands. 
The system can still develop a charge excitation gap at nonzero integer filling without breaking any valley or spin symmetries. 
This is also what we observed in our numerical calculations. Additionally, we find that no other nodes exist at other momentum points. 
On the other hand, if \cref{eq:gap_condition} is not satisfied, the system must be gapless since the middle two bands are degenerate at $\Gamma$ point.

\subsubsection{Comparison of the Hubbard-I spectrum in two limits}
We now compare the Hubbard-I spectrum (at $M=0,v_\star^\prime=0$) in two limits: $U_1 < \gamma$ and $U_1 > \gamma$. Our focus is on the Fermi velocity of the gapless node at the $\Gamma$ point for $\nu = 0$. 
At $\nu=0$, we have $\epsilon_f = 0, \nu_f=0,epsilon_c=0$,. By diagonalizing the auxiliary Hamiltonian \cref{eq:aux_ham_flat_chiral_limit} within the Hubbard-I approximation, we obtain the following two eigenvalues that describe a Dirac node
\begin{align}
    \pm \sqrt{ 
    \frac{\gamma^2 + |v_\star \kk|^2 +(\frac{U_1}{2})^2}{2}- \sqrt{ 
    \bigg[ \frac{\gamma^2 + |v_\star \kk|^2 + (\frac{U_1}{2})^2}{2} \bigg]^2-\frac{U_1^2 |v_\star \kk|^2}{4}
    } 
    } 
\end{align}
A small momentum expansion leads to 
\begin{align}
    \pm \sqrt{ \frac{ U_1^2}{4\gamma^2 +U_1^2 }} |v_\star \kk| 
\end{align}
with velocity $v_{D} = \sqrt{\frac{U_1^2}{4\gamma^2 +U_1^2} } v_\star $. We could then get the velocity in two limits which are 
\begin{align}
    U_1\gg |\gamma|&:\quad v_D \approx v_\star \nonumber\\ 
    U_1 \ll |\gamma| &: \quad v_D \approx \frac{U_1}{2|\gamma|}v_\star 
\end{align}
In the limit of $ U_1 \gg |\gamma|$, the fermion velocity approaches that of graphene. As $ U_1$ decreases, the fermion velocity becomes gradually renormalized. 
Taking the parameters derived from ab-initio and given in Ref.~\cite{song_magic-angle_2022}, we observe $\sqrt{\frac{U_1^2}{4\gamma^2 +U_1^2} }\approx  0.76$. Then the velocity of Dirac dispersion is $v_D \approx 3.27\text{eV}\cdot \text{\AA}$. However, we also comment that the effective Hubbard $U_1$ (which characterizes the separation of Hubbard bands) could be further renormalized due to the finite hybridization as shown in the DMFT simulation \cite{rai_dynamical_2023}. This could further renormalize the velocity of Dirac dispersions.

\subsection{Effective coupling between flavor moments} 
In the limit of $|\gamma|/|v_\star \Lambda_c| \rightarrow 0$, we demonstrated that the system can be described by atomic action given in~\cref{eq:eff_action_atomic_limit}. A small but non-zero $|\gamma|/|v_\star \Lambda_c|$ will generate coupling between moments formed by the $f$ electrons. 
In this section, we estimate the effective coupling in the projected limit. Since the flat bands are primarily composed of $f$ electrons with strong Hubbard interactions, we can apply a mean-field decoupling to the Hubbard repulsion term for the $f$ electrons.
\begin{align}
    H_{U_1} \approx& -\frac{U_1}{2}\sum_{\RR} \nu_f^2 + U_1 \nu_f \sum_{\RR,\alpha \eta s}:n_{\RR,\alpha\eta s}^{f}: \nonumber\\ 
    &
    +\frac{U_1}{2}\sum_{\RR,\alpha\eta s,\alpha'\eta's'} O_{\alpha\eta s, \alpha'\eta's'}(\RR) O_{\alpha'\eta' s', \alpha\eta s}(\RR) -U_1 \sum_{\RR,\alpha\eta s,\alpha'\eta's'}O_{\alpha\eta s, \alpha'\eta's'}(\RR):f_{\RR,\alpha'\eta's'}^\dag f_{\RR,\alpha\eta s}: 
\end{align}
where 
\begin{align}
    &O_{\alpha\eta s,\alpha'\eta's'}(\RR) = \langle :f_{\RR,\alpha\eta s}^\dag f_{\RR,\alpha'\eta's'}:\rangle \nonumber\\ 
    &\nu_f = \frac{1}{N}\sum_\RR \text{Tr}[O(\RR)]
\end{align}
and we have assumed the system does not develop CDW phase with a non-uniform charge distribution of $f$ electrons. We can observe that $O_{\alpha\eta s,\alpha'\eta's'}(\RR)$ describes the spin, valley, and orbital moments (flavor moments) formed by the $f$ electrons. 

In terms of the path-integral, the corresponding action reads
\begin{align}
 S = &S_0 \nonumber\\ 
    & 
    +\int_\tau \bigg[\frac{U_1}{2}\sum_{\RR,\alpha\eta s,\alpha'\eta's'} O_{\alpha\eta s, \alpha'\eta's'}(\RR,
\tau) O_{\alpha'\eta' s', \alpha\eta s}(\RR,\tau) -U_1 \sum_{\RR,\alpha\eta s,\alpha'\eta's'}O_{\alpha\eta s, \alpha'\eta's'}(\RR,\tau):f_{\RR,\alpha'\eta's'}^\dag(\tau) f_{\RR,\alpha\eta s}(\tau): 
    \bigg] 
\end{align}
where $S_0$ contains the single-particle term of the $f$ and $c$ electrons and can be written as 
\begin{align}
    S_0 =&\frac{1}{\beta}\sum_{i\omega,\kk, \alpha \eta s,\alpha'\eta's' } 
    \begin{bmatrix}
        f_{\kk,\alpha\eta s}^\dag(i\omega) & c_{\kk,\alpha \eta s}^\dag(i\omega) & c_{\kk,\alpha +3 \eta s}^\dag(i\omega)
    \end{bmatrix} \nonumber\\ 
    &
    \begin{bmatrix}
       i\omega-  \epsilon_f & \gamma & 0\\
       \gamma & i\omega -\epsilon_c & v_\star (\eta k_x +i (-1)^{\alpha+1}k_y) \\ 
       0&  v_\star (\eta k_x +i (-1)^{\alpha+1}k_y) & i\omega -\epsilon_c
    \end{bmatrix} \begin{bmatrix}
        f_{\kk,\alpha\eta s}(i\omega) \\ c_{\kk,\alpha \eta s}(i\omega)\\  c_{\kk,\alpha +3 \eta s}(i\omega)
    \end{bmatrix}-\frac{U_1}{2}\sum_\RR \nu_f^2 
\end{align}
Here, $\epsilon_f,\epsilon_c$ includes the Hartree contributions of $\hat{H}_{U_1},\hat{H}_W,\hat{H}_V$ (see Ref.~\cite{hu_kondo_2023}).

We can then integrate out the $c$ and $f$ electrons which leads to the following action 
\begin{align}
    S_{eff} = &
    \int_\tau \bigg[ \frac{U_1}{2}\sum_{\RR,\alpha\eta s,\alpha'\eta's'} O_{\alpha\eta s, \alpha'\eta's'}(\RR) O_{\alpha'\eta' s', \alpha\eta s}(\RR)\bigg]  -\text{Tr}\log\bigg\{ [ G_{0}^{-1} ]
    -V[O]]
    \bigg] 
    \bigg\} 
    \label{eq:eff_action}
\end{align}
where $G_0$ is the Green's function 
\begin{align}
    [G_0]_{\RR \tau \alpha\eta s, \RR' \tau' \alpha'\eta's'} = -\langle T_\tau  f_{\RR,\alpha\eta s}(\tau) f^\dag_{\RR',\alpha'\eta's'}(\tau')\rangle 
\end{align}
calculated with respective to the action $S_0$, 
and the vertex that describes the coupling between $O(\RR)$ fields and electrons is defined as 
\begin{align}
    \bigg[ V[O]\bigg]_{\RR \tau \alpha\eta s, \RR' \tau' \alpha'\eta's'} = \delta_{\RR',\RR"}\delta_{\tau,\tau'} O_{\alpha\eta s,\alpha'\eta's'}(\RR)U_1
\end{align}
$S_{eff}$ can be understood as the effective action of the fields $O(\RR)$. From $S_{eff}$, we can derive the coupling between flavor moments ($O(\RR)$) of different sites. 

We begin by discussing the properties of the Green's function. Since we are focusing only on the contribution from the flat band, we could project the $f$ electron operator into the band basis (~\cref{eq:d-operator})
\begin{align}
    f^\dag_{\kk,\alpha\eta s}(\tau) = \frac{1}{\sqrt{\mathcal{N}_\kk} }d_{\kk, (-1)^{\alpha+1} ,\eta s}^\dag +...
\end{align}
where $...$ represents the contribution from the remote bands. To simplify the problem, we only include the contribution of the flat bands, since flat bands are near the Fermi energy. 
Then Green's function could be written as
\begin{align}
\label{eq:green_fun}
  [G_0]_{\RR \tau \alpha\eta s, \RR' \tau' \alpha'\eta's'}=&\frac{1}{N_M}\sum_\kk 
  \langle -T_\tau f_{\kk,\alpha\eta s}(\tau) f^\dag_{\kk,\alpha'\eta
  s'}(\tau') \rangle e^{i\kk\cdot(\RR-\RR')}
  \end{align} 
Approximately, we find 
\begin{align}
    [G_0]_{\RR \tau \alpha\eta s, \RR' \tau' \alpha'\eta's'}
  \approx &
  \frac{\delta_{\alpha\eta s, \alpha'\eta's'}}{N_M}\sum_\kk \frac{1}{{\mathcal{N}_\kk} }
  \langle -T_\tau d_{\kk, (-1)^{\alpha+1} ,\eta s}(\tau) d^\dag_{\kk, (-1)^{\alpha+1} ,\eta s}(\tau')  \rangle e^{i\kk\cdot(\RR-\RR')}
\end{align}
where only the contributions from the flat band have been included. 

To get a simple analytical expression, we assume the flat band is perfectly flat, meaning that $\langle d_{\kk, (-1)^{\alpha+1} ,\eta s}(\tau) d^\dag_{\kk, (-1)^{\alpha+1} ,\eta s}(\tau') \rangle $ is $\kk$-independent.
Under this assumption, we can introduce the following Green's functions 
\begin{align}
    &g_0(\tau-\tau') = -\langle T_\tau d_{\kk, (-1)^{\alpha+1} ,\eta s}(\tau) d^\dag_{\kk, (-1)^{\alpha+1} ,\eta s}(\tau') \rangle  ,\quad g_0(i\omega) = \int_0^\beta g_0(\tau) d\tau 
\end{align} 
For perfect flat bands at Fermi energy, the Green's function takes the simple form of $g_0(i\omega)=1/i\omega$. Then we have 
\begin{align}
[G_0]_{\RR \tau \alpha\eta s, \RR' \tau' \alpha'\eta's'}=&\frac{1}{\beta}\sum_{i\omega} 
\frac{1}{i\omega} e^{-i\omega (\tau-\tau')}\frac{1}{N_M}\sum_\kk \frac{1}{\mathcal{N}_\kk} e^{i\kk\cdot(\RR-\RR')}
\delta_{\alpha\eta s,\alpha'\eta's'}
\end{align}
where the $\RR$-dependency of Green's function arises from the wavefunction contribution, characterized by $1/\sqrt{\mathcal{N}_\kk}$. For further convenience, we introduce 
\begin{align}
  I(\RR-\RR') =  \frac{1}{N_M}\sum_\kk \frac{1}{\mathcal{N}_\kk} e^{i\kk\cdot(\RR-\RR')} =\frac{1}{N_M}\sum_\kk \frac{1}{1+\gamma^2/|v_\star \kk|^2} e^{i\kk\cdot(\RR-\RR')}
\end{align}
and then 
\begin{align}
    [G_0]_{\RR \tau \alpha\eta s, \RR' \tau' \alpha'\eta's'}=&\frac{1}{\beta}\sum_{i\omega} 
\frac{1}{i\omega} e^{-i\omega (\tau-\tau')}I(\RR-\RR')\delta_{\alpha\eta s,\alpha'\eta's'}
\end{align}

We are now in the position to evaluate the effective action defined in~\cref{eq:eff_action}. 
We begin by separating the Green's function into local and nonlocal contribution
\begin{align}
 &[G_0]_{\RR \tau \alpha\eta s, \RR' \tau' \alpha'\eta's'} =  [G_{loc}]_{\RR \tau \alpha\eta s, \RR'\tau' \alpha'\eta's'} + [\delta G]_{\tau \alpha\eta s, \tau' \alpha'\eta's'}  \nonumber\\ 
 &[G_{loc}]_{\RR \tau \alpha\eta s, \RR'\tau' \alpha'\eta's'} = \delta_{\RR,\RR'}[G_0]_{\RR \tau \alpha\eta s, \RR' \tau' \alpha'\eta's'} \nonumber\\ 
 &[\delta G]_{\tau \alpha\eta s, \tau' \alpha'\eta's'} = (1-\delta_{\RR,\RR'})[G_0]_{\RR \tau \alpha\eta s, \RR' \tau' \alpha'\eta's'} 
\end{align}
$G_{loc}$ and $\delta G$ correspond to the on-site (called here local) $(\RR=\RR')$ and off-site (called here nonlocal) $(\RR\ne \RR')$ contribution of Green's function. The nonlocal Green's function couples the flavor-moment fields $O(\RR)$ between different sites. 
 
We then treat the nonlocal Green's function $\delta G$ as a perturbation and derive the effective coupling of the flavor moments $O(\RR)$. With $G_{loc}, \delta G$, we can rewrite $S_{eff}$ as 
\begin{align}
    S_{eff} = &
    \int_\tau \bigg[ \frac{U_1}{2}\sum_{\RR,\alpha\eta s,\alpha'\eta's'} O_{\alpha\eta s, \alpha'\eta's'}(\RR,
\tau) O_{\alpha'\eta' s', \alpha\eta s}(\RR,\tau)\bigg]  -\text{Tr}\log\bigg\{ [G_{loc} +\delta G]^{-1}
    -V[O]     \bigg\}  \nonumber\\ 
    = &
    \int_\tau \bigg[ \frac{U_1}{2}\sum_{\RR,\alpha\eta s,\alpha'\eta's'} O_{\alpha\eta s, \alpha'\eta's'}(\RR,
\tau) O_{\alpha'\eta' s', \alpha\eta s}(\RR,\tau)\bigg] \nonumber\\ 
    & -\text{Tr}\log\bigg\{ [G_{loc} +\delta G]^{-1} \bigg\} 
    -\text{Tr}\log\bigg\{ \mathbb{I} - [G_{loc} +\delta G]
    V[O] 
    \bigg\} 
\end{align}

At zeroth order in $\delta G$, we have an atomic action 
\begin{align}
    S_{atom} 
    \approx &
    \int_\tau \bigg[\frac{U_1}{2}\sum_{\RR,\alpha\eta s,\alpha'\eta's'} O_{\alpha\eta s, \alpha'\eta's'}(\RR,
\tau) O_{\alpha'\eta' s', \alpha\eta s}(\RR,\tau)\bigg]-\text{Tr}\log\bigg\{ [G_{loc}]^{-1}
    -V[O]
    \bigg\} 
\end{align}
where each site is decoupled.

We can perform a Taylor expansion in powers of $\delta G$ which gives 
\begin{align}
    &-\text{Tr}\log\bigg\{ \mathbb{I} - [G_{loc} +\delta G]
    V[O] 
    \bigg\}  \nonumber\\ 
    =& \sum_{n=0} \frac{1}{n}\text{Tr}\bigg\{\bigg[(G_{loc} +\delta G)
    V[O] \bigg]^n \bigg\}\nonumber\\ 
    \approx  &  \sum_{n=0} \frac{1}{n}\text{Tr}\bigg\{\bigg[G_{loc}
    V[O] \bigg]^n \bigg\} + \sum_{n=2}^{\infty}\sum_{k=0}^{n-2} \text{Tr}
    \bigg[\delta G V[O] \bigg(G_{loc}V[O]\bigg)^{k} \delta G V[O] 
    \bigg(G_{loc}V[O]\bigg)^{n-2-k} 
    \bigg]\,.
\end{align}
The leading-order coupling between flavor moments generated by the nonlocal Green's function $\delta G$ is 
\begin{align}
\label{eq:eff_action_spin_spin_coupling}
    S' = &  \sum_{n=2}^{\infty}\sum_{k=0}^{n-2} \text{Tr}
    \bigg[\delta G V[O] \bigg(G_{loc}V[O]\bigg)^{k} \delta G V[O] 
    \bigg(G_{loc}V[O]\bigg)^{n-2-k} 
    \bigg]\ ,.
\end{align}
We now evaluate $S'$ explicitly. We first notice that the local Green's function is
\begin{align}
\label{eq:local-green-mat}
   [G_{loc}]_{\RR \tau \alpha\eta s,\RR' \tau'
    \alpha'
    \eta'
    s'}  = \frac{1}{\beta}\sum_{i\omega}\delta_{\alpha\eta s,\alpha'\eta's'}\delta_{\RR,\RR'}\frac{I(\bm{0})}{i\omega-\epsilon_0} e^{-i\omega (\tau-\tau')}
\end{align}
and the nonlocal Green's function is 
\begin{align}
\label{eq:non-local-green-mat}
     [\delta G]_{\RR \tau \alpha\eta s,\RR' \tau'
    \alpha'
    \eta'
    s'}  = \frac{1}{\beta}\sum_{i\omega}\delta_{\alpha\eta s,\alpha'\eta's'}(1-\delta_{\RR,\RR'})\frac{I(\RR-\RR')}{i\omega-\epsilon_0} e^{-i\omega (\tau-\tau')}
\end{align}
As for the $V[O]$ term, we can introduce a $SU(8)$ rotation matrix $T(\RR)$ and a diagonal matrix $S$ such that 
\begin{align}
\label{eq:decomp_spin_field}
    O(\RR) = T(\RR)S T^\dag(\RR)
\end{align} 
At integer filling, we approximately have $\nu_f \approx \nu \in \mathbb{Z}$. We can represent the diagonal matrix as
\begin{align}
\label{eq:def-diag-comp-O-field}
    &S = \text{diag}\{ \frac{1}{2},\frac{1}{2},\frac{1}{2},\frac{1}{2}, -\frac{1}{2},-\frac{1}{2},-\frac{1}{2},-\frac{1}{2}\},\quad \nu=0\nonumber\\ 
    &S = \text{diag}\{ \frac{1}{2},\frac{1}{2},\frac{1}{2},-\frac{1}{2}, -\frac{1}{2},-\frac{1}{2},-\frac{1}{2},-\frac{1}{2}\},\quad \nu=-1\nonumber\\ 
     &S = \text{diag}\{ \frac{1}{2},\frac{1}{2},-\frac{1}{2},-\frac{1}{2}, -\frac{1}{2},-\frac{1}{2},-\frac{1}{2},-\frac{1}{2}\},\quad \nu=-2\nonumber\\ 
     &S = \text{diag}\{ \frac{1}{2},-\frac{1}{2},-\frac{1}{2},-\frac{1}{2}, -\frac{1}{2},-\frac{1}{2},-\frac{1}{2},-\frac{1}{2}\},\quad \nu=-3
\end{align}
The diagonal components of $S$ with a value of $1/2$ correspond to the flavors that are filled, while those with a value of $-1/2$ correspond to the flavors that are empty. This gives $\text{Tr}[O(\RR)]=\text{Tr}[S]=\nu$. 
The decomposition in~\cref{eq:decomp_spin_field} gives 
\begin{align}
\label{eq:int-vertex-mat}
\bigg[V[O]\bigg]_{\RR \tau \alpha\eta s,\RR' \tau'
    \alpha'
    \eta'
    s'} = U_1\delta_{\RR,\RR"}\delta_{\tau-\tau'}  \bigg[ T(\RR)S T^\dag(\RR)\bigg]_{\alpha\eta s,\alpha'\eta's'}
\end{align}
Combining~\cref{eq:eff_action_spin_spin_coupling,eq:local-green-mat,eq:non-local-green-mat,eq:int-vertex-mat}, we find 
\begin{align}
    S' =\frac{1}{\beta}\sum_{i\omega ,\RR,\RR'} \sum_{n=2}^{\infty}\sum_{k=0}^{n-2} \frac{I(\RR-\RR')I(\RR'-\RR) I(\bm{0})^{n-2}}{(i\omega -\epsilon_0)^n}U_1^{n}\text{Tr}
    \bigg[ T_{\RR'}S^{1+k} T_{\RR'}^\dag T_{\RR}S^{n-1-k}T^\dag_{\RR}
    \bigg]
    \label{eq:eff_action_spin_spin_coupling_itermed}
\end{align}
Since $S$ is a diagonal matrix (\cref{eq:def-diag-comp-O-field}), we find
\begin{align}
    &S^n = \frac{1}{2^{n-1}} S,\quad n \in 2\mathbb{Z}+1\nonumber\\ 
    &S^n = \frac{1}{2^{n-1}}\mathbb{I},\quad n \in 2\mathbb{Z}
\end{align}
This indicates that 
\begin{align}
    &\text{Tr}
    \bigg[ T_{\RR'}S^{n_1} T_{\RR'}^\dag T_{\RR}S^{n_2}T^\dag_{\RR}
    \bigg] = \frac{1}{2^{n_1+n_2-2}}8,\quad n_1 \in 2\mathbb{Z},n_2 \in 2\mathbb{Z} \nonumber\\ 
    &
    \text{Tr}
    \bigg[ T_{\RR'}S^{n_1} T_{\RR'}^\dag T_{\RR}S^{n_2}T^\dag_{\RR}
    \bigg] = \frac{1}{2^{n_1+n_2-1}}\text{Tr}[T_{\RR'}ST_{\RR'}^\dag] = \frac{1}{2^{n_1+n_2-1}}\text{Tr}[O({\RR'})]=\frac{\nu_f}{2^{n_1+n_2-1}},\quad n_1 \in 2\mathbb{Z}+1,n_2 \in 2\mathbb{Z} \nonumber\\ 
    &
    \text{Tr}
    \bigg[ T_{\RR'}S^{n_1} T_{\RR'}^\dag T_{\RR}S^{n_2}T^\dag_{\RR}
    \bigg] =\frac{\nu_f}{2^{n_1+n_2-1}},\quad n_1 \in 2\mathbb{Z},n_2 \in 2\mathbb{Z} +1\nonumber\\  
    &
    \text{Tr}
    \bigg[ T_{\RR'}S^{n_1} T_{\RR'}^\dag T_{\RR}S^{n_2}T^\dag_{\RR}
    \bigg] =\frac{1}{2^{n_1+n_2-2}}\text{Tr}[T_{\RR'}ST_{\RR'}^\dag T_{\RR'}ST_{\RR'}^\dag]= \frac{1}{2^{n_1+n_2-2}}\text{Tr}[O(\RR)O(\RR')]
    ,\quad n_1 \in 2\mathbb{Z}+1,n_2 \in 2\mathbb{Z} +1
\end{align}
 Only terms with $n_1\in 2\mathbb{Z}+1,n_2\in 2\mathbb{Z}+1$ generate a coupling between moments ($O(\RR)$) between different sites. Therefore, we focus on the terms with $1+k\in 2\mathbb{Z}+1, n-1-k \in 2\mathbb{Z} +1 $ in ~\cref{eq:eff_action_spin_spin_coupling_itermed}. This also implies that $n \in 2\mathbb{Z}, k \in 2\mathbb{Z}$. We let $n=2m,k=2s$, and the corresponding contributions are
\begin{align}
    S'_{eff} =&\frac{1}{\beta}\sum_{i\omega ,\RR,\RR'} \sum_{m=1}^{\infty}\sum_{s=0}^{m-1} \frac{I(\RR-\RR')I(\RR'-\RR) I(\bm{0})^{2m-2}}{(i\omega -\epsilon_0)^n}U_1^{2m}\text{Tr}
    \bigg[ T_{\RR'}S^{1+2s} T_{\RR'}^\dag T_{\RR}S^{2m-1-2s}T^\dag_{\RR}
    \bigg] \nonumber\\ 
    =&
    \frac{1}{\beta}\sum_{i\omega ,\RR,\RR'} \sum_{m=1}^{\infty}\sum_{s=0}^{m-1} \frac{I(\RR-\RR')I(\RR'-\RR) I(\bm{0})^{2m-2}}{(i\omega -\epsilon_0)^{2m} 2^{2m-2}}U_1^{2m}\text{Tr}
    \bigg[ O(\RR) O(\RR')
    \bigg] \nonumber\\ 
    =&   \frac{1}{\beta}\sum_{i\omega ,\RR,\RR'} I(\RR-\RR')I(\RR'-\RR)\text{Tr}
    \bigg[ O(\RR) O(\RR')
    \bigg] \frac{(i\omega)^2 U^2}{\bigg[\omega^2 + \bigg(\frac{I(\bm{0})U}{2}\bigg)^2\bigg]^2} \nonumber\\ 
    =&-\beta \sum_{\RR,\RR'}I(\RR-\RR')I(\RR'-\RR) \frac{U}{4\pi I(\bm{0})} \text{Tr}[O(\RR)O(\RR')] \nonumber\\ 
    =& \beta\sum_{\RR,\RR'}J(\RR-\RR')\text{Tr}[O(\RR)O(\RR')]
\end{align}
In the last line, we have introduced the following coupling between the flavor moments
\begin{align}
J(\RR-\RR') = -I(\RR-\RR')I(\RR'-\RR) \frac{U}{4\pi I(\bm{0})} =  -|I(\RR-\RR')|^2 \frac{U}{4\pi I(\bm{0})}
\end{align} 
Several comments are in order 
\begin{itemize}
    \item $J(\RR-\RR') $ is always ferromagnetic, which indicates the system always tends to develop ferromagnetic order in the limit we considered. We comment that a similar ferromagnetic coupling has also been derived in the unprojected limit of THF in Ref.~\cite{hu_kondo_2023}. 
    \item We could refine our estimation of the effective spin-spin coupling by numerically evaluating Green's function in~\cref{eq:green_fun}, which will include the contributions from the dispersions of flat bands and the remote bands. 
    \item The coupling we derived is $U(8)$ symmetric. Additional symmetry-breaking terms in the THF model, which were ignored in the current calculation--such as the $v_\star^\prime, M$--could generate effective couplings that break this $U(8)$ symmetry. 
\end{itemize}

We can estimate the ordering temperatures of the ferromagnetic phase. The energy gained by developing a ferromagnetic order ($\qq=0$ order) is proportional to
\begin{align}
    J(\qq=0) = \sum_{\RR \ne \bm{0}}J(\RR) 
\end{align}
Here, we have subtracted $\RR=0$ contribution, as the $\RR=0$-term corresponds to a local coupling of the form $\text[O(\RR)O(\RR)]$ and does not favor any specific types of order. 
Written explicitly, we have 
\begin{align} 
\label{eq:def_spin_spin_coupl_q0}
    J_{ferro} =J(\qq=0)=  &\sum_{\RR }J(\RR)  -J(\bm{0})= -\sum_\RR |I(\RR)|^2 \frac{U}{4\pi I(\bm{0})} + \frac{U}{4\pi }I(0)  \nonumber\\ 
    =& -\frac{1}{N_M^2}\sum_{\kk_1,\kk_2}\sum_{\RR}\frac{U}{4\pi I(\bm{0})}\frac{1}{\mathcal{N}_{\kk_1}\mathcal{N}_{\kk_2}} e^{i(\kk_1-\kk_2)\RR}  + \frac{U}{4\pi }I(0) \nonumber\\
    = &- \frac{1}{N_M}\sum_{\kk_1,\kk_2}\frac{U}{4\pi}\frac{\delta_{\kk_1,\kk_2}}{\mathcal{N}_{\kk_1}\mathcal{N}_{\kk_2}I(\bm{0})} +\frac{U}{4\pi} I(0)
    \end{align}
We then replace the $\kk$ summation with a momentum integral
    \begin{align}
      \frac{1}{N_M}\sum_\kk  \approx \frac{1}{\Omega_{M}}\int_{|\kk|<\Lambda_c}
\end{align}
We can approximately take $\Omega_{M} \approx \pi \Lambda_c^2$ as we have discussed below~\cref{eq:approx_sum_to_int}. The momentum integral gives
\begin{align}
\label{eq:1_NkNk_integral}
    \frac{1}{N_M} \sum_{\kk_1}\frac{1}{\mathcal{N}_{\kk}^2} \approx & \frac{1}{\Omega_{M}}\int_{|\kk|<\Lambda_c}  \frac{1}{\bigg[1+\gamma^2/|v_\star \kk|^2\bigg]^2} 
    \approx  1 + \frac{\gamma^2}{\Lambda_c^2|v_\star|^2}\bigg( 1 - 
    \log\frac{\gamma^4}{v_\star^4\Lambda^4}\bigg)
\end{align}
where we have also expanded the final results in powers of $\gamma^2/|\Lambda v_\star|^2$. 
In addition, we have
\begin{align}
\label{eq:I0_integral}
    I(\bm{0}) = \frac{1}{N_M}\sum_\kk \frac{1}{\mathcal{N}_\kk} \approx \frac{1}{\Omega_{M}}\int_{|\kk|<\Lambda_c} \frac{1}{1+\gamma^2/|v_\star \kk|^2}= 1-\frac{\gamma^2}{v_\star^2 \Lambda_c^2}\log(1+ \frac{v_\star^2\Lambda^2}{\gamma^2} )
\end{align}
Combining~\cref{eq:def_spin_spin_coupl_q0,eq:1_NkNk_integral,eq:I0_integral}, we obtain
\begin{align}
   J_{\rm ferro} 
    \approx&-\frac{U}{\pi}\frac{\gamma^2}{v_\star^2 \Lambda_c^2} \log\bigg( \frac{ |\Lambda_c v_\star|}{|\gamma|}\bigg)
\end{align}
We can then conclude the temperature scale for the development of ferromagnetic order is 
$ T_{\rm ferro} \sim \frac{U}{\pi}\frac{\gamma^2}{v_\star^2 \Lambda_c^2} \log\bigg( \frac{ |\Lambda_c v_\star|}{|\gamma|}\bigg)
$. 
We then expect the quasi-free flavor moments to appear within the temperature window $ U_1 \gtrsim  T \gtrsim T_{\rm ferro}$.
Finally, we comment that this ferromagnetic coupling persists in the regime where $U_1$ is larger than the gap between the remote band and flat band~\cite{hu_kondo_2023}, allowing for continuity between the projected and the unprojected descriptions. 

\section{Comparison between topological heavy-fermion model and non-local-moment model}
\label{sec:comp}
In this section, we provide a detailed comparison between our results obtained from the THF model and those presented in Ref.~\cite{ledwith_nonlocal_2024}. 
As summarized in~\cref{tab:comp_table}, the results in Ref.~\cite{ledwith_nonlocal_2024} are consistent with our results. This consistency suggests that the THF model is applicable for studying both the strong coupling limit with $U_1 > |\gamma|$ and the projected limit with $U_1<|\gamma|$. Furthermore, 
it suggests that the ``non-local-moment behavior" discussed in Ref.~\cite{ledwith_nonlocal_2024} could be understood via the Hubbard-I approximation of the THF model.

\section{Quantitative aspects of the THF model}
\label{sec:THF-parameter}

In this section, we discuss how the parameters of THF compare with experimental observations. In particular, $U_1$ is consistent with the cascade peaks in STM experiments as well as band gaps at fillings $\nu=\pm4$. However, the agreement of other parameters with experiments remains less certain.

\begin{figure}[th]
\centering
\includegraphics[width=1\linewidth]{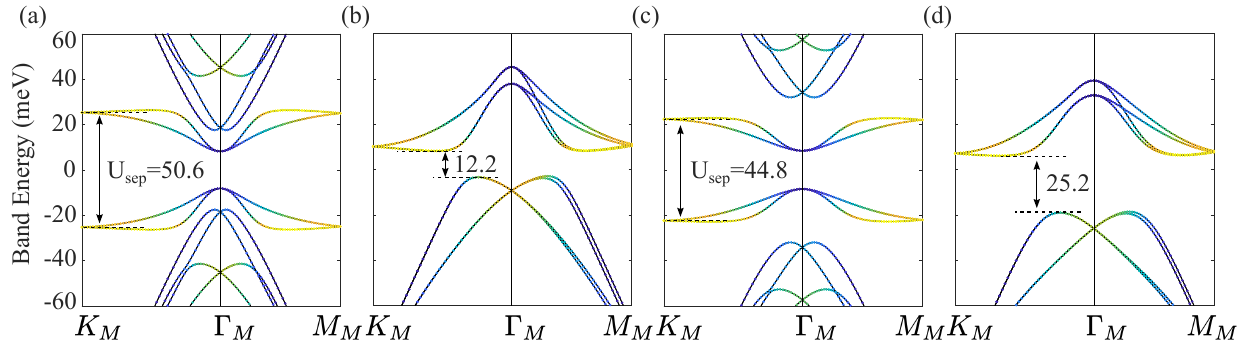}
\caption{Hartree-Fock bands of the THF model. 
    (a) and (c) are Hartree-Fock bands at CNP with parameters $w_0/w_1=0.8, 0.7$, respectively. 
    (b) and (d) are Hartree-Fock bands at $\nu=-4$ with parameters $w_0/w_1=0.8$, 0.7, respectively. 
    The $U_1=$58.0, 51.7meV for  $w_0/w_1=0.8$, 0.7, respectively.
}
\label{fig:HF-bands}
\end{figure}

We adopt parameters at $\theta=1.05^\circ$ and $w_0/w_1=$0.8, 0.7.
The corresponding parameters of the THF model, as presented in Ref.~\cite{song_magic-angle_2022} and Ref.~\cite{calugaru_tbg_2023}, are derived from the microscopic model, with a dielectric constant $\epsilon = 6$ and a screening length $\xi =10$nm. 
Variations in the screening length and dielectric constant can alter these parameters. 
These parameters have also been used to study the local-moment behaviors in the DMFT calculations \cite{hu_kondo_2023,zhou_kondo_2024,rai_dynamical_2023}. 
Here we plot the Hartree-Fock band structures (\cref{fig:HF-bands}(a), (c)) at CNP with a Kramers inter-valley coherent order assumed. 
(As explained in Ref.~\cite{song_magic-angle_2022}, the value of the gap does not depend on the order much.)
The separation ${U}_{\rm sep}$ between two $f$-bands is found 50.6meV and 44.8meV for $w_0/w_1=$0.8 and 0.7, respectively.
The DMFT ${U}_{\rm sep}$ for a symmetric state at CNP is about 40meV for $w_0/w_1=$0.8 \cite{rai_dynamical_2023}. 
${U}_{\rm sep}$ is smaller than the {\it bare} $U_1$ parameter (58meV and 52meV) because of the $cf$ hybridization; and it is ${U}_{\rm sep}$ rather than $U_1$ should be compared to experiments. 
For comparison, we summarize ${U}_{\rm sep}$ seen in STM experiments of samples featuring quantum-dot-like cascades of transitions \cite{wong_cascade_2020,choi_interaction-driven_2021,oh_evidence_2021,nuckolls_quantum_2023} that motivated the THF model.
From Fig.~1d of Ref.~\cite{wong_cascade_2020}, Fig.~4a of Ref.~\cite{choi_interaction-driven_2021}, Fig.~4g of Ref.~\cite{choi_interaction-driven_2021}, Fig.~1c of Ref.~\cite{oh_evidence_2021}, Fig.~3a of Ref.~\cite{nuckolls_quantum_2023}, one can extract $U_{\rm sep}\approx$45meV, 50meV, 49meV, 53meV, 40meV, respectively. 
As such, the theoretical $U_{\rm sep}$'s quantitatively match experimental values.
Had it not matched the experiments, it would suggest an issue in the understanding of the parameters, since the values $U_1$ are \emph{not} postulated but microscopically derived based on ab initio calculations and dielectric constants quoted in the experiment.

In addition, the compressibility obtained from the DMFT (Fig.~7a of Ref.~\cite{rai_dynamical_2023}) and Gutzwiller \cite{hu_link_2024} approaches both produce four cascades features from $\nu=0$ to $\nu=-4$, which is also consistent with experimental observation (Fig.~7a of Ref.~\cite{rai_dynamical_2023}). The changes of chemical potential from $\nu=0$ to $\nu= \pm 4$ obtained from DMFT calculations also match the experimental observations as shown in Fig.~7b of Ref.~\cite{rai_dynamical_2023}. 

We also calculate and plot the symmetric Hartree-Fock bands at $\nu=-4$ in \cref{fig:HF-bands}(b), (d).  
The band gap is found 12.2meV and 25.2meV for $w_0/w_1=$0.8 and 0.7, respectively, consistent with the measured gap (15meV) by SET experiment \cite{zondiner_cascade_2020}.

Finally, we provide further discussions on the limit of 
\begin{align}
    \gamma/|v_\star \Lambda_c| \rightarrow 0 \, . 
\end{align}
In this limit, the Berry curvature becomes highly concentrated and is effectively described by a $\delta$-function (\cref{eq:berry_curv_d_op_simp}). However, such a curvature can only be realized in a system with band touching. This implies that the gap between the remote bands and the dispersive bands vanishes, making it necessarily smaller than any finite $U_1$, which itself must be larger than the bandwidth of the narrow bands. 
This can be shown on general grounds (see \cref{sec:berry_curvature}) and can also be observed within the THF model. If we fix the bandwidth of the $c$-electrons ($|v_\star \Lambda_c|$), then taking the limit $\gamma / |v_\star \Lambda_c| \to 0 $ necessarily leads to \( \gamma \to 0 \), indicating that the gap between the remote and flat bands disappears in the heavy-fermion model. 
Thus, in this limit, due to band touching, one cannot simply focus on (or project onto) the flat bands while neglecting the remote bands. However, the heavy-fermion model, which explicitly incorporates both flat and remote bands, remains a valid framework for studying both regimes.

\section{Summary}
\label{sec:summary}
In this work, we study the THF model of MATBG in the projected limit of 
\begin{align}
\label{eq:weak_hyb_limit_2}
    |v_\star \Lambda_c| \gg |\gamma| \gg U_1
\end{align}
We demonstrate that the THF model remains a faithful description of the system in the projected limit (\cref{eq:weak_hyb_limit_2}), allowing for a straightforward analysis of the topology and correlations. 
In fact, the THF model was originally inspired by calculations in the momentum-space model within the projected limit~\cite{BER21a,LIA21,BER21b}, where the single-particle excitation 
exhibits flat dispersion, indicating the presence of heavy $f$ electrons. 
Through the THF model, we have derived the flat-band wavefunction (also present in Ref.~\cite{song_magic-angle_2022}), the self-energy of the flat-bands, and the effective coupling between flavor moments. 



\begin{acknowledgments}
We thank D. C\u{a}lug\u{a}ru, T. Wehling, G. Sangiovanni, R. Valenti, P. J. Ledwith, and E. Khalaf for useful discussions.
H. H. was supported by the European Research Council (ERC) under the European Union’s Horizon 2020 research and innovation program (Grant Agreement No. 101020833).
H. H. was also supported by the Gordon and Betty Moore Foundation through Grant No.GBMF8685 towards the Princeton theory program, the Gordon and Betty Moore Foundation’s EPiQS Initiative (Grant No. GBMF11070), Office of Naval Research (ONR Grant No. N00014-20-1-2303), Global Collaborative Network Grant at Princeton University, BSF Israel US foundation No. 2018226, NSF-MERSEC (Grant No. MERSEC DMR 2011750), Simons Collaboration on New Frontiers in Superconductivity and the Schmidt Foundation at the Princeton University.
Z.-D. S. was supported by National Natural Science
Foundation of China (General Program No. 12274005), National Key Research and Development Program of China (No. 2021YFA1401900), and Innovation Program for Quantum Science and Technology (No. 2021ZD0302403).
B. A. B was supported by DOE Grant No. DESC0016239. 
\end{acknowledgments}

\section{Appendix: Mapping an interacting Green's function to a non-interacting Green's function}
\label{app:sec:hubbar_I:mapp_green_fun}
In this appendix, we demonstrate that the Green's function of an interacting system can be equivalently described by the Green's function of non-interacting systems with additional auxiliary fermions. To show it, we start from a generic multi-orbital system with electron operators $\gamma_{i}$ where $i$ can denote momentum, orbital, sublattice indices. The system can be described by the following generic interacting Hamiltonian 
\begin{eqnarray}
    \hat{H} = \sum_{ij}t_{ij}\gamma_i^\dag \gamma_j + \sum_{ijlm}U_{ijlm}\gamma_i^\dag \gamma_j \gamma_l^\dag \gamma_m 
\end{eqnarray}
where $t_{ij}$ is the hopping matrix, and $U_{ijlm}$ describes the interactions. 

The Green's function is then 
\begin{align}
    &G_{ij}(\tau) = -\langle T_\tau \gamma_i(\tau) \gamma^\dag_j(0)\rangle \nonumber\\ 
    &G_{ij}(i\omega) = \int_0^\beta G_{ij}(\tau)e^{i\omega_n\tau}d\tau 
    = 
\bigg[ 
        i\omega \mathbb{I} - t - \Sigma(i\omega) 
\bigg]_{ij}^{-1}
\label{app:eq:hubbard_i_int_green}
\end{align}
where $\Sigma_{ij}(i\omega) $ is the self-energy.

\subsection{Properties of Green's function}
We first discuss the properties of the Green's function. 
For a generic electron operator $\gamma_{i}$ with flavor index $i$ (momentum, orbital, spin, sublattice, etc.). We define the Green's function as 
\begin{align}
    G_{ij}(\tau) = -\langle T_\tau \gamma_i(\tau) \gamma^\dag_j(0)\rangle 
\end{align}
We assume the eigenstates and eigenenergy of the Hamiltonians are $|n\rangle, E_n$ respectively with 
\begin{align}
    H|n\rangle = E_n |n\rangle 
\end{align}
Then from the definition of Green's function we have  (for $\beta >\tau \ge 0$)
\begin{align}
    G_{ij}(\tau) = -\frac{1}{Z} \sum_{n,m} e^{-(\beta-\tau)E_n-\tau E_m}\langle n| \gamma_i |m\rangle \langle m|\gamma_j^\dag |n\rangle ,\quad Z= \sum_n e^{-\beta E_n}
\end{align}
Performing Fourier transformation, we could obtain the Green's function in the Matsubara frequency
\begin{align}
    G_{ij}(i\omega) = \int_0^\beta G_{ij}(\tau) e^{i\omega_n\tau} d\tau = \frac{1}{Z} \sum_{n,m}\frac{1}{i\omega - E_m+E_n}
    \bigg(e^{-\beta E_m} + e^{-\beta E_n} 
    \bigg)\langle n| \gamma_i |m\rangle \langle m|\gamma_j^\dag |n\rangle
\end{align}
By performing analytical continuation, the Green's function on the entire complex plane reads 
\begin{align}
    G_{ij}(z) = \int_0^\beta G_{ij}(\tau) e^{i\omega_n\tau} d\tau = \frac{1}{Z} \sum_{n,m}\frac{1}{z- E_m+E_n}
    \bigg(e^{-\beta E_m} + e^{-\beta E_n} 
    \bigg)\langle n| \gamma_i |m\rangle \langle m|\gamma_j^\dag |n\rangle
\end{align}
The asymptotic behaviors of Green's function at large $|z|$ is
\begin{align}
    G_{ij}(z) \sim &\frac{1}{Z}\sum_{n,m} \langle n|\gamma_i |m\rangle \langle m|\gamma_j^\dag |n\rangle \bigg(e^{-\beta E_m} + e^{-\beta E_n} 
    \bigg)\frac{1}{z}\bigg[1+\frac{E_m-E_n}{z} +O(1/z^2)\bigg]\nonumber\\
    =&\sum_n \frac{\langle n|\gamma_i \gamma_j^\dag +\gamma_j^\dag \gamma_i |n\rangle e^{-\beta E_n}}{Z}\frac{1}{z} +\frac{N_{ij}}{z^2} +O(1/z^3)
\nonumber\\&
    =\frac{ \delta_{i,j}}{z}  +\frac{N_{ij}}{z^2}+ O(1/z^3)
\end{align}
where 
\begin{align}
    N_{ij} = \frac{1}{Z}\sum_{n,m}\sum_{n,m} \langle n|\gamma_i |m\rangle \langle m|\gamma_j^\dag |n\rangle (e^{-\beta E_m}+e^{-\beta E_n}) (E_m-E_n)
\end{align}
We can also observe that $N_{ij} = N_{ji}^*$ indicating $N$ is an Hermitian matrix.

The spectral functions are 
\begin{align}
    \rho_{ij}(\omega) = \frac{1}{\pi}\text{Im}[G_{ij}(\omega-i0^+)] = 
    \frac{1}{Z} \sum_{n,m}\delta(\omega -E_m+E_n)
    \bigg(e^{-\beta E_m} + e^{-\beta E_n} 
    \bigg)\langle n| \gamma_i |m\rangle \langle m|\gamma_j^\dag |n\rangle
\end{align}
From the definition of the spectral function, we can observe that it is Hermitian with
\begin{align}
    [\rho_{ij}(\omega)]^* = \rho_{ji}(\omega)
\end{align}
In addition, we also find
\begin{align}
    G_{ij}(z) = \int_{-\infty}^{\infty} \frac{\rho_{ij}(\omega)}{z-\omega}d\omega
\end{align}
which corresponds to the spectral representation of Green's function. 
The spectral function also follows the sum rule 
\begin{align}
    &\int_{-\infty}^{\infty}\rho_{ij}(\omega) 
    = \frac{1}{Z}\sum_{n,m}  
    \bigg(e^{-\beta E_m} + e^{-\beta E_n} 
    \bigg)\langle n| \gamma_i |m\rangle \langle m|\gamma_j^\dag |n\rangle \nonumber\\
    =& \frac{1}{Z}
    \bigg[\sum_m e^{-\beta E_m}\langle m| \gamma_j^\dag \gamma_i|m\rangle  
    + \sum_n e^{-\beta E_n} \langle n|\gamma_i \gamma_j^\dag|n\rangle 
    \bigg] =\frac{1}{Z}\sum_m 
    \bigg[e^{-\beta E_m} 
    \langle m|\gamma_j^\dag \gamma_i +\gamma_i^\dag\gamma_j |m\rangle \bigg] \nonumber\\
    =& \delta_{i,j}\frac{1}{Z}\sum_me^{-\beta E_m} \nonumber\\
    =&\delta_{i,j}
\end{align}

\subsection{Properties of self-energy}
In this section, we discuss the properties of the self-energy. From the definition, it behaves as 
\begin{align}
    \Sigma_{ij}(z) = z\delta_{i,j} - [H_0]_{ij}- G_{ij}^{-1}(z)
\end{align}
In the large $|z| \gg 1 $ limit we have 
\begin{align}
    \Sigma_{ij}(z) \sim z\delta_{i,j} - [H_0]_{ij} - z\delta_{i,j} +N_{ij} + O(1/z^2)\sim -[H_0]_{ij}+N_{ij} +O(1/z)
\end{align}
Therefore $\Sigma_{ij}(z)$ may not decay to zero at $|z|\rightarrow \infty$. However, one could separate the constant contributions from the self-energy and let 
\begin{align}
    \Sigma_{ij}(z) = \delta \Sigma_{ij}(z) + N_{ij}-[H_0]_{ij}
\end{align}
where $\delta\Sigma_{ij}(z)$ is an analytical function and decay as $1/|z|$. Therefore, $\delta \Sigma(z)$ follows Kramer-Kronig relations.

\subsection{Mapping an interacting Green's function to a non-interacting Green's function}
\label{app:sec:hubbar_I:mapp_green_fun}
In this appendix, we demonstrate that the Green's function of an interacting system can be equivalently described by the Green's function of non-interacting systems with additional auxiliary fermions. To show it, we start from a generic multi-orbital system with electron operators $\gamma_{i}$ where $i$ can denote momentum, orbital, sublattice indices. The system can be described by the following generic interacting Hamiltonian 
\begin{eqnarray}
    \hat{H} = \sum_{ij}t_{ij}\gamma_i^\dag \gamma_j + \sum_{ijlm}U_{ijlm}\gamma_i^\dag \gamma_j \gamma_l^\dag \gamma_m 
\end{eqnarray}
where $t_{ij}$ is the hopping matrix, and $U_{ijlm}$ describes the interactions. 

The Green's function is then 
\begin{align}
    &G_{ij}(\tau) = -\langle T_\tau \gamma_i(\tau) \gamma^\dag_j(0)\rangle \nonumber\\ 
    &G_{ij}(i\omega) = \int_0^\beta G_{ij}(\tau)e^{i\omega_n\tau}d\tau 
    = 
\bigg[ 
        i\omega \mathbb{I} - t - \Sigma(i\omega) 
\bigg]_{ij}^{-1}
\label{app:eq:hubbard_i_int_green}
\end{align}
where $\Sigma_{ij}(i\omega) $ is the self-energy. 
We could separate the self-energy into two parts
\begin{align}
    \Sigma(i\omega) = \delta \Sigma(i\omega) + N
\end{align}
with $N = \lim_{|\omega|\rightarrow\infty} \Sigma(\omega)$. 
We can utilize the spectral decomposition of the self-energy which reads
\begin{align}
    \delta \Sigma_{ij}(i\omega_n) = \int_{-\infty}^{\infty} \frac{ \rho^{\Sigma}_{ij}(\epsilon) }{i\omega_n - \epsilon } d\epsilon 
\end{align}
where $\rho^{\sigma}_{ij}(\epsilon)$ is the spectral functions of the self-energy functions and can be calculated from the real-frequency self-energy 
\begin{align}
\rho_{ij}^{\Sigma}(\omega) = -\frac{1}{2\pi i }\left[\delta \Sigma_{ij}(\omega+i0^+) -\delta \Sigma_{ji}^*(\omega+i0^+) \right]
\label{app:eq:hubbard_i_spec_self_energy}
\end{align}
We note that, from \cref{app:eq:hubbard_i_spec_self_energy}, the following properties of the self-energy holds
\begin{align}
    \rho_{ij}^{\Sigma}(\omega) =\bigg[ \rho_{ji}^{\Sigma}(\omega)\bigg]^*
\end{align}

We then approximate the spectral functions $\rho_{ij}^{\Sigma}(\omega)$ via a series of $\delta(x)$ functions as 
\ba 
\rho_{ij}^{\Sigma}(\omega) \approx \sum_n v_i^n (v_j^n)^* \delta(\omega- \epsilon_n)
\ea 
where $v_{i}^n, \epsilon_n$ are introduced to describe the behaviors of $\rho_{ij}^{\Sigma}$.
This also indicates that the self-energy can be written as
\begin{align}
\label{app:eq:hubbard_i_self_energy}
   \delta  \Sigma_{ij}(i\omega) = \sum_{n} \frac{v_{i}^n v_j^{n,*}}{i\omega-\epsilon_n }
\end{align}

We now prove that the interacting Green's function (\cref{app:eq:hubbard_i_int_green}) is equivalent to the Green's function of a non-interacting system.

We first introduce the following non-interacting systems
\begin{equation}
    \hat{H}^{aux} = \sum_{ij}\tilde{t}_{ij}\gamma_i^\dag \gamma_j + \sum_{n}\epsilon^n  a_{n}^\dag a_{n} + \sum_{n} 
    \bigg[v_{i}^n \gamma_{i}^\dag a_{n}+\text{h.c.}
    \bigg]  
    \label{eq:app:aux_ham}
\end{equation} 
where $\tilde{t} = t -N$ which contains both the non-interacting contributions and the static (non-decay) contributions from the self-energy.  
We can now calculate the corresponding Green's functions with respect to the Hamiltonian $\hat{H}^{aux}$. The Green's function (which is a matrix) is then
\ba 
G^{aux}(\tau) = 
\langle T_\tau 
\begin{bmatrix}
    \gamma(\tau) & a(\tau) 
\end{bmatrix}\cdot 
\begin{bmatrix}
    \gamma^\dag (0) \\ a^\dag(0) 
\end{bmatrix} \rangle 
\ea 
where $\gamma(\tau) = [\gamma_1(\tau),...]$ is the vector formed by all the $\gamma$ electron operators, and $a(\tau) = [a_{1}^\dag(\tau),...]$ is the vector formed by all the $a$ electron operators. The Green's function then can be calculated as
\ba 
G^{aux}(i\omega_n) 
=
\begin{bmatrix}
  i\omega_n \mathbb{I} - \tilde{t} & -{V} \\
-V^\dag  & i\omega_n \mathbb{I} - \epsilon 
\end{bmatrix}^{-1}
\ea 
where the hybridization matrix is 
\ba 
V_{i, n} = v_i^n 
\ea 
 and $\epsilon$ is a diagonal matrix with diagonal elements $\epsilon_n$. 
We now aim to get the Green's function of $\gamma$. We first note the following fact 
\begin{equation}
   \begin{bmatrix}
       A & B \\ 
       C & D 
   \end{bmatrix}^{-1} = 
    \begin{bmatrix}
        (A-BD^{-1}C)^{-1} & -(A-BD^{-1}C)^{-1}BD^{-1} \\ 
        -(D-CA^{-1}B)^{-1} CA^{-1} & (D-CA^{-1}B)^{-1}
    \end{bmatrix}
\end{equation}
with $A,B,D,C$ matrices. Then we can calculate the Green's function of $\gamma$ electrons by taking the upper left block of the $G^{aux}(i\omega_n)$
\begin{align} 
\label{app:eq:hubbard_i_int_green_2}
&G^{aux,\gamma}_{ij}(i\omega_n)= 
\langle T_\tau \gamma_{i}(\tau)\gamma^\dag_j(0)\rangle  
=
\begin{bmatrix}
    i\omega_n\mathbb{I} -\tilde{t} - \tilde{\Sigma}(i\omega_n)
\end{bmatrix}^{-1}_{ij}
\end{align} 
where we have introduced 
\begin{align}
\label{eq:app:eq:hubbard_i_self_energy_2}
&\tilde{\Sigma}_{ij}(i\omega_n) = \bigg[ V\cdot [i\omega_n\mathbb{I}-\epsilon]^{-1} \cdot V\bigg]_{ij}  = \sum_n \frac{v_{i}^n (v_{j}^n)^*}{i\omega_n-\epsilon_{n}}
\end{align}
By comparing \cref{app:eq:hubbard_i_self_energy} and \cref{eq:app:eq:hubbard_i_self_energy_2}, we can conclude that 
\begin{equation}
    \tilde{\Sigma}_{ij}(i\omega_n) =  \delta\Sigma_{ij}(i\omega_n)
\end{equation}
Therefore, via \cref{app:eq:hubbard_i_int_green} and \cref{app:eq:hubbard_i_int_green_2}, we conclude that
\begin{equation}
    G_{ij}^{aux,\gamma}(i\omega_n) = G_{ij}(i\omega_n)
\end{equation}
Then, we can conclude that, the interacting Green's function can be equivalently described by the Green's function of an effective non-interacting system. However, we note that, the mapping itself will not help us to calculate the interacting Green's function, but allows us to more conveniently study the band topology of the interacting system via the effective non-interacting system.

\section{Appendix: Berry curvature and band gap in a two-orbital toy model}
\label{sec:berry_curvature}
In this section, we analyze a simple two-orbital model to explore the relationship between the band gap and Berry curvature. We derive the Berry curvature at a general momentum $\kk$ and show that a singular Berry curvature can only emerge when the band gap closes. Additionally, we examine the Berry curvature near the saddle point of the band and demonstrate that, as the gap closes, the Berry curvature can approach a $\delta$-function behavior near the saddle point.

We assume the hopping matrix of the two-orbital model (in two dimensions) takes the form of \begin{align}
    H_\kk = \sum_{\mu=x,y,z} d_\mu(\kk) \sigma^\mu 
\end{align}
where $\sigma^{x,y,z}$ are Pauli matrices, and $d_{x,y,z}(\kk)$ are three real functions of $\kk$. 
The dispersion reads 
\begin{align}
   E_{\kk,\pm}= \pm \sqrt{\sum_\mu [d_\mu(\kk)]^2 }
\end{align} 
We assume the system is fully gapped with $\sqrt{\sum_\mu [d_\mu(\kk)]^2} \ne 0$ for all $\kk$. 

We now discuss the direct gap and the Berry curvature. 
For a generic $\kk$ point $\kk_0$, we could perform a small $k\cdot p$ expansion with 
\begin{align}
\label{eq:k_p_expansion}
    H_{\kk_0 + (p_x,p_y)} = H_{\kk_0}  + p_x D_x [\kk_0]+ p_y D_y[\kk_0]
\end{align}
where $D_{x,y}[\kk_0] $ are two $2\times 2$ Hermitian and tracless matrices (since $H_\kk$ is traceless). 
We then perform a basis transformation denoted by $U(\kk_0)$ such that
\begin{align}
    U(\kk_0)^\dag H_{\kk_0}U(\kk_0) = E_{\kk_0,+} \sigma_z 
\end{align}
We can further introduce 
\begin{align}
    h_x[\kk_0] = U(\kk_0)^\dag D_x[k_0]U(\kk_0),\quad 
     h_y[\kk_0] = U(\kk_0)^\dag D_y[k_0]U(\kk_0)
\end{align}
Then the Hamiltonian near $\kk_0$ can be written as 
\begin{align}
    U(\kk_0)^\dag H_{\kk_0 + (p_x,p_y)}U(\kk_0) \approx 
    E_{\kk_0} \sigma_z  + h_x[\kk_0] p_x + h_y[\kk_0]p_y
\end{align}
In addition, we parametrize the $h_x[\kk_0], h_y[\kk_0]$ as 
\begin{align}
    h_x[\kk_0]= v_x[\kk_0] [\mathbf{n_x}[\kk_0]\cdot \bm{\sigma}],\quad 
    h_y[\kk_0]= v_y[\kk_0] [\mathbf{n_y}[\kk_0]\cdot \bm{\sigma}]
\end{align}
where $\mathbf{n_x}[\kk_0],\mathbf{n_y}[\kk_0]$ are two unit vectors, and 
\begin{align}
\label{eq:def_v_matrix}
    v_{x/y}[\kk_0] = \sqrt{ \frac{1}{2}\text{Tr}[h_{x/y}[\kk_0]\cdot h_{x/y}[\kk_0]] }
\end{align}
We now evaluate the Berry curvature. It is sufficient to consider the lowest band. The wavefunction can be written as 
\begin{align}
& u[\kk_0+(p_x,p_y)] = 
    \frac{1}{\sqrt{2\sqrt{X^2+Y^2+Z^2}(\sqrt{X^2+Y^2+Z^2}-Z})}
    \begin{bmatrix}
        Z-\sqrt{X^2+Y^2+Z^2} & X+iY
    \end{bmatrix},\quad \nonumber\\ 
    &X = \sum_{\mu=x,y}p_\mu v_\mu[\kk_0]n_\mu^x 
    ,\quad X =  \sum_{\mu=x,y}p_\mu v_\mu[\kk_0]n_\mu^y,\quad Z=  \sum_{\mu=x,y}p_\mu v_\mu[\kk_0]n_\mu^z + E_{\kk_0}
\end{align}
The Berry curvature can be calculated as 
\begin{align}
    \Omega(\kk_0+(p_x,p_y)) = i\bigg( \partial_x u[\kk_0+(p_x,p_y)]^*
    \partial_y u[\kk_0+(p_x,p_y)] -  \partial_y u[\kk_0+(p_x,p_y)]^*
    \partial_x u[\kk_0+(p_x,p_y)] \bigg) 
\end{align}
Exactly at $\kk_0$, we observe the Berry curvature takes the form of 
\begin{align}
\label{eq:berry_curvature_generic_k}
    \Omega(\kk_0) = \frac{1}{2}\frac{v_x[\kk_0]v_y[\kk_0]}{|E_{\kk_0,+}|^2} (\mathbf{n}_x[\kk_0]\times \mathbf{n}_y[\kk_0])\cdot \mathbf{e}_z
\end{align}
Here, we could observe that $\Omega(\kk_0) $ is proportional to $1/|E_{\kk_0,+}|^2$. To obtain a diverged Berry curvature, we need to either let $|v_x[\kk_0]v_y[\kk_0| \rightarrow \infty$, or let $|E_{\kk_0,+}|^2\rightarrow 0$. 
However, we notice that (from \cref{eq:parameterization,eq:def_v_matrix,eq:k_p_expansion}),
\begin{align}
     v_\mu[\kk_0]^2= \frac{1}{2} \text{Tr}\bigg[ [\partial_{\mu}H_{\kk_0}]^2  \bigg] 
\end{align}
For a generic matrix with 
\begin{align}
    H_{\kk} = \sum_{n,m,\mu}t_{n,m}^\mu \sigma^\mu  e^{-i\kk\cdot(n\mathbf{a}_1+m\mathbf{a}_2)}
\end{align}
where $t_{n,m}^\mu$ is the corresponding hopping strength, $\mathbf{a}_1,\mathbf{a}_2$ are lattice vectors. We have 
\begin{align}
    [ v_\mu[\kk_0]]^2 = \sum_\nu 
   \bigg| \sum_{n,m}t_{n,m}^\nu (n\mathbf{a}_1 + m\mathbf{a}_2)_\mu\bigg|^2 
\end{align}
For an exponentially decayed hopping model, taking $|v_\mu[\kk_0]|^2\rightarrow \infty $ is thus infeasible. Therefore, the only feasible way to realize a singular Berry curvature is to reduce the direct gap $|E_{\kk_0,+}|$ of the system at the corresponding momentum point. This then suggests a band-touching point.

Since the Berry curvature is proportional to $1/|E_{\kk_0,+}|^2$, it is also useful to investigate its behaviors near the saddle point, at which $1/|E_{\kk_0,+}|^2$ could reach its maximum/minimum. We consider a generic saddle point where 
\begin{align}
    \partial_{k^\mu} E_{\kk_0,\pm} = 0
    \label{eq:saddle_point_disp}
\end{align}
Since 
\begin{align}
    &E_{\kk_0+(p_x,p_y), \pm } \nonumber\\ 
    \approx &\pm  \sqrt{ (E_{\kk_0,+}+ v_x[\kk_0]{n}^x_z[\kk_0]p_x+v_y[\kk_0]n_y^z[\kk_0]p_y)^2 +
    ( v_x[\kk_0]{n}^x_x[\kk_0]p_x+v_y[\kk_0]n_y^x[\kk_0]p_y)^2 
    +( v_x[\kk_0]{n}^y_x[\kk_0]p_x+v_y[\kk_0]n_y^y[\kk_0]p_y)^2 
    } \nonumber\\
    \approx & \pm (E_{\kk_0,+}+ v_x[\kk_0]{n}^x_z[\kk_0]p_x+v_y[\kk_0]n_y^z[\kk_0]p_y) + O(p_x^2,p_y^2,p_{x}p_y)
\end{align}
\cref{eq:saddle_point_disp} requires 
\begin{align}
  v_x[\kk_0]  \mathbf{n}_x[\kk_0]\cdot \mathbf{e}_z =  v_y[\kk_0]\mathbf{n}_y[\kk_0]\cdot \mathbf{e}_z = 0 
\end{align}
This can be realized by taking $v_{x/y}[\kk_0]=0$ or $\mathbf{n}_{x/y}[\kk_0]\cdot\mathbf{e}_z=0$. 
If either $v_{x}[\kk_0]=0$ or $v_y[\kk_0]=0$, we have 
\begin{align}
    \Omega_{\kk_0} = 0 
\end{align}
which corresponds to a trivial saddle point. 
For the other situations with $\mathbf{n}_{x}[\kk_0]\cdot\mathbf{e}_z=0$ and $\mathbf{n}_{x}[\kk_0]\cdot\mathbf{e}_z=0$, we find 
\begin{align}
\label{eq:berry_near_k_0}
    \Omega(\kk_0 + (p_x,p_y))=  \frac{v_x[\kk_0]v_y[\kk_0]  |E_{\kk_0,+}|
   ( \mathbf{n}_x[\kk_0]\times \mathbf{n}_y[\kk_0])\cdot\mathbf{e}_z
    }{2\bigg[|E_{\kk_0,+}|^2 +( v_x[\kk_0]p_x)^2 
    +( v_y[\kk_0]p_y)^2 
    +2 v_x[\kk_0]p_xv_y[\kk_0]p_y (\mathbf{n}_{x}[\kk_0]\cdot 
    \mathbf{n}_{y}[\kk_0])
    \bigg] 
    ^{3/2}}
\end{align}
We could observe that 
\begin{align}
    \partial_{p_\mu} \Omega(\kk_0+(p_x,p_y))\bigg|_{p_x=p_y=0} = 0 
\end{align}
which indicates that, the saddle point of the band dispersion also corresponds to the saddle point of the Berry curvature. 
The second-order derivative of the Berry curvature behaves as 
\begin{align}
   &  \partial_{p_x}^2 \Omega(\kk_0+(p_x,p_y))\bigg|_{p_x=p_y=0} = 
     -\frac{3 ( \mathbf{n}_x[\kk_0]\times \mathbf{n}_y[\kk_0])\cdot\mathbf{e}_z}{2 |E_{\kk_0,+}|^{4} }
   (  v_x[\kk_0])^3 v_y[\kk_0] \nonumber\\ 
   &
     \partial_{p_y}^2 \Omega(\kk_0+(p_x,p_y))\bigg|_{p_x=p_y=0} = 
     -\frac{3 ( \mathbf{n}_x[\kk_0]\times \mathbf{n}_y[\kk_0])\cdot\mathbf{e}_z}{2 |E_{\kk_0,+}|^{4} }
   (  v_y[\kk_0])^3 v_x[\kk_0] \nonumber\\ 
   &  \partial_{p_x}\partial_{p_y} \Omega(\kk_0+(p_x,p_y))\bigg|_{p_x=p_y=0} = 
     \frac{3 (\mathbf{n}_x[\kk_0]\cdot \mathbf{n}_y[\kk_0]) ( \mathbf{n}_x[\kk_0]\times \mathbf{n}_y[\kk_0])\cdot\mathbf{e}_z}{2 |E_{\kk_0,+}|^{4} }
   (  v_x[\kk_0])^2 (v_y[\kk_0])^2
\end{align}
All the eigenvalues of matrix $M_{\mu\nu} =\partial_{p_\mu}\partial_{p_\nu} \Omega(\kk_0+(p_x,p_y)))|_{p_x=p_y=0} $ have the same sign. This suggests that the Berry curvature reaches the local minimum/maximum correspondingly at the saddle point. 

We next discuss the behaviors of the Berry curvature near $\kk_0$. We could let 
\begin{align}
\label{eq:parameterization}
    &p_{x} = p\cos(\theta),\quad p_y = p\sin(\theta)  \nonumber\\ 
    &v_x[\kk_0]= V\cos(\theta_v),\quad v_y[\kk_0]= V\sin(\theta_v)\nonumber\\
    & ( \mathbf{n}_x[\kk_0]\times \mathbf{n}_y[\kk_0])\cdot\mathbf{e}_z = \sin(\phi) ,\quad  \mathbf{n}_x[\kk_0]\cdot \mathbf{n}_y[\kk_0] = \cos(\phi)
\end{align}
From \cref{eq:berry_near_k_0}, we find 
\begin{align}
    & \Omega(\kk_0+(p_x,p_y)) =  \frac{V^2 |E_{\kk_0,+}|\sin(\phi)\sin(2\theta_v)}{4 
     \bigg[ |E_{\kk_0,+}|^2 + 
     V^2p^2 f(\theta,\theta_v,\phi)
     \bigg]^{3/2}
     }\nonumber\\ 
     &f(\theta,\theta_v,\phi) = \cos(\theta)^2 \cos(\theta_v)^2 
     +\sin(\theta)^2\sin(\theta_v)^2 +\frac{1}{2}\sin(2\theta_v)\sin(2\theta)\cos(\phi)
\end{align}
We could observe that as we take $|E_{\kk_0}|\rightarrow 0$ which indicates a band touching point, the Berry curvature distribution becomes a $\delta$ function.

\bibliography{tbg}

\end{document}